\documentclass[sigplan,10pt,authorversion]{acmart}
\usepackage{graphicx}
\graphicspath{{./images/}}
\usepackage{algorithm}
\usepackage{booktabs}
\usepackage{url}
\usepackage{multirow}
\usepackage[noend]{algpseudocode}
\usepackage{array}
\usepackage{subcaption}
\usepackage{comment}
\usepackage{enumitem}
\usepackage{tikz}
\usepackage{cleveref}

\makeatletter

\renewcommand\subsection{\@startsection{subsection}{2}{\z@}%
  {-.75\baselineskip \@plus -2\p@ \@minus -.2\p@}%
  {.25\baselineskip}%
  {\ACM@NRadjust\@subsecfont}}

\makeatother

\copyrightyear{2021}
\acmYear{2021}
\setcopyright{acmlicensed}\acmConference[SOSP '21]{ACM SIGOPS 28th Symposium on Operating Systems Principles}{October 26--29, 2021}{Virtual Event, Germany}
\acmBooktitle{ACM SIGOPS 28th Symposium on Operating Systems Principles (SOSP '21), October 26--29, 2021, Virtual Event, Germany}
\acmPrice{15.00}
\acmDOI{10.1145/3477132.3483551}
\acmISBN{978-1-4503-8709-5/21/10}

\crefname{theorem}{Theorem}{Theorems}
\crefname{definition}{Definition}{Definitions}
\crefname{proposition}{Proposition}{Propositions}
\crefname{lemma}{Lemma}{Lemmas}
\crefname{claim}{Claim}{Claims}
\crefname{corollary}{Corollary}{Corollaries}
\crefname{property}{Property}{Properties}
\crefname{algorithm}{Algorithm}{Algorithms}
\crefname{listing}{Listing}{Listings}
\crefname{section}{\S}{\S\S}
\crefname{appendix}{Appendix}{Appendices}
\crefname{figure}{Figure}{Figures}
\crefname{table}{Table}{Tables}

\newcommand*\circled[1]{\tikz[baseline=(char.base)]{
            \node[shape=circle,draw,inner sep=.7pt] (char) {#1};}}

\begin{document}

\title{Cuckoo Trie: Exploiting Memory-Level Parallelism for Efficient DRAM Indexing}

\author{Adar Zeitak}
\affiliation{
  \institution{Tel Aviv University, Israel}
  \country{}
}

\author{Adam Morrison}
\affiliation{
  \institution{Tel Aviv University, Israel}
  \country{}
}

\begin{abstract}
We present the Cuckoo Trie, a fast, memory-efficient ordered index structure.
The Cuckoo Trie is designed to have memory-level parallelism---which a modern out-of-order processor
can exploit to execute DRAM accesses in parallel---without sacrificing memory efficiency.
The Cuckoo Trie thus breaks a fundamental
performance barrier faced by current indexes, whose bottleneck is a series of dependent pointer-chasing
DRAM accesses---e.g., traversing a search tree path---which the processor cannot parallelize.
Our evaluation shows that the Cuckoo Trie outperforms state-of-the-art-indexes by up to 20\%--360\%
on a variety of datasets and workloads, typically with a smaller or comparable memory footprint.
\end{abstract}

\maketitle

\newtheorem*{unnumbered_definition}{Definition}

\providecommand{\floor}[1]{\left \lfloor #1 \right \rfloor}

\section{Introduction}

Modern enterprise and cloud systems increasingly rely on \emph{in-memory} database systems~\cite{Silo,Hekaton,H-Store,redis,HyPer,SAP-HANA}.
In-memory systems store the entire dataset in main memory, avoiding the overhead of slow block-based storage media, and thereby achieving
high query throughput and low latency.

Databases rely on \emph{ordered indexes} to accelerate query processing. An ordered index is a dictionary data structure
that supports ordered iteration over the stored keys in addition to standard point queries, such as key lookups.
In-memory systems crucially require fast and memory-efficient indexes, as indexing accounts
for a significant fraction of query execution time (as high as 94\%~\cite{MeetTheWalkers}) and database memory consumption (as much as 55\%~\cite{hybrid-indexes,OracleBlog}).

Consequently, much effort has been invested in adapting the design of classic indexes---such as B-trees~\cite{UB-Tree}, tries~\cite{trie}, and skip lists~\cite{skiplist}---to modern hardware architectures. These indexes all share a common structure: the index
is a hierarchical graph in memory (e.g., a tree), such that traversing through it gradually ``shrinks'' the reachable
key space until finally the target key is found (or proven missing). Modern indexes thus focus on shortening the searched paths to
reduce the number of memory accesses~\cite{art,hot}, reducing per-node work performed by the traversal~\cite{hot,k-ary,fast,parallel-skiplist}, and cache-friendly memory layouts of the structure~\cite{csb,css-tree,hydra-list,masstree}.

\begin{figure}
	\includegraphics[width=0.95\linewidth]{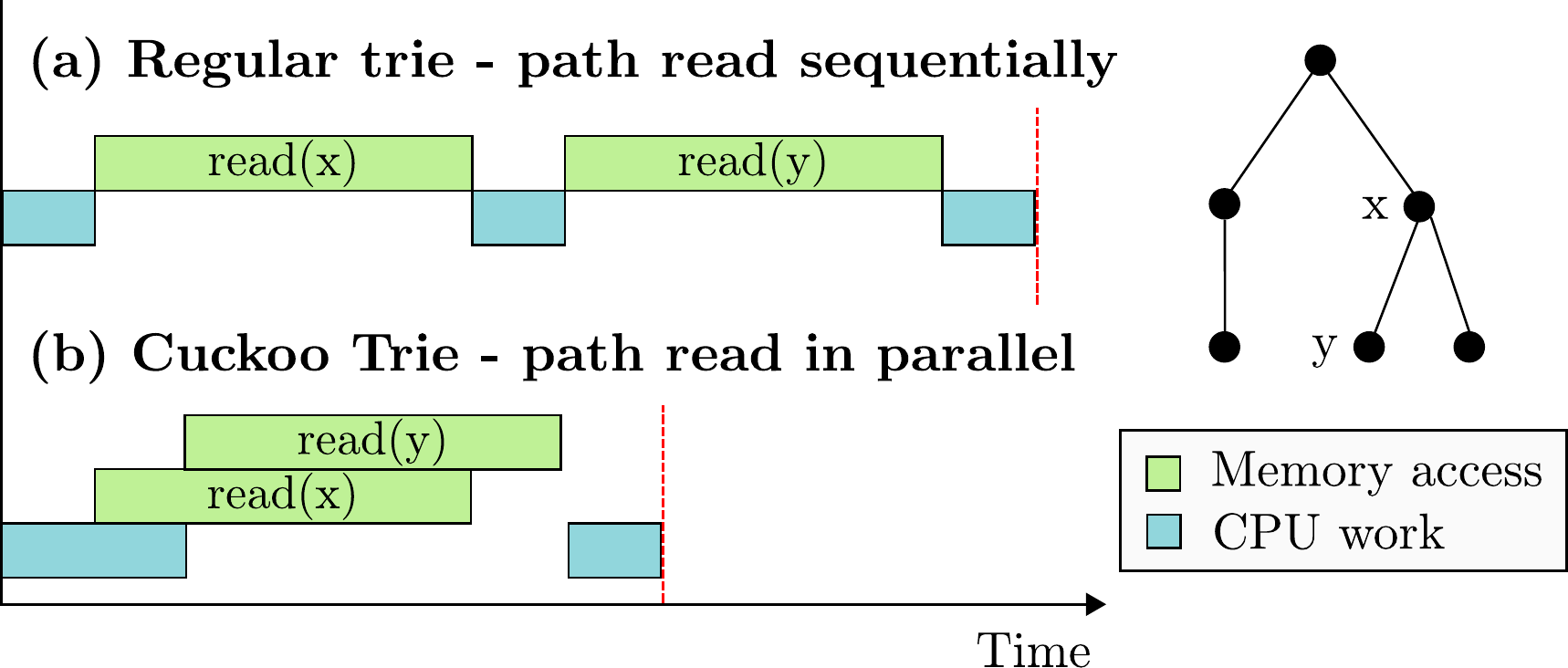}
    \vspace{-10pt}
	\caption{CPU and memory activity during a trie lookup. The Cuckoo Trie's MLP-aware design allows the CPU to read a child ($y$) in parallel with its parent ($x$).}
    \vspace{-10pt}
	\label{fig:mlp}
\end{figure}

But these indexes squander a key performance feature of modern processors: exploiting \emph{memory-level parallelism} (MLP)
by executing multiple independent memory reads in parallel, to overlap their latency~\cite{mlp-def}.
Because the underlying index structure is a pointer-based graph, its traversal is \emph{inherently sequential}:
the address of the next node on a path is read from the current node.
This creates a significant bottleneck, as the size of modern-day datasets requires most nodes to be fetched from DRAM~\cite{MeetTheWalkers}
and DRAM access latency is equivalent to the execution time of hundreds of instructions~\cite{MemoryWall}.
The processor thus sits mostly idle, waiting for memory reads to complete (\cref{fig:mlp}a).
The result is a fundamental performance barrier for current index designs.

We present the \emph{Cuckoo Trie} to break this barrier. The Cuckoo Trie is a \emph{high-MLP, memory-efficient} ordered index.
In the Cuckoo Trie, MLP is a primary design goal. MLP is achieved by replacing the pointer-based trie with an implicit \emph{hashed representation} that stores trie nodes in a Cuckoo hash table, keyed by the node ``names'' (prefixes of the keys).
This representation allows any node on the root-to-key path to be read directly, without having to read its parent first,
as the key's prefixes are known up front (\cref{fig:mlp}b).

The Cuckoo Trie's key goal is to be fast, scalable, and general-purpose \emph{without} trading memory efficiency in exchange for
these properties. In contrast, the only MLP-aware index we know of~\cite{mlp-index}
has $3\times$ the memory footprint of the Cuckoo Trie; inherently supports only 8-byte keys; and does not support range scans
nor concurrency.

A major challenge addressed by the Cuckoo Trie is achieving memory efficiency.
Normally, a hash table stores copies of its keys---in our case, trie key prefixes---which imposes significant
memory overhead. The Cuckoo Trie solves this problem using a novel hash table design that \emph{eliminates} key storage
(direct or indirect) from hash table nodes.

We evaluate the Cuckoo Trie on a variety of workloads and datasets, including when integrated into Redis, a popular in-memory data store.
In both single- and multi-threaded runs, the Cuckoo Trie outperforms the state-of-the-art indexes HOT~\cite{hot}, ART~\cite{art}, Wormhole~\cite{wormhole}, and STX~\cite{stx} by
up to 35\%, 20\%, 50\%, and 360\%, respectively, typically with reduced (up to 28\%) or similar memory usage.
Overall, the Cuckoo Trie offers a better speed vs. memory efficiency trade-off than prior indexes: none of them is both
smaller and faster than it.

\noindentparagraph{Contributions.}
We make the following contributions:
\begin{itemize}[leftmargin=*]
\item \textbf{Problem:} Identifying lack of MLP as a fundamental barrier to index performance (\cref{sec:need-mlp}).
\item \textbf{Cuckoo Trie:} A high-MLP memory-efficient ordered index, based on a novel
hash table design that \emph{eliminates} key storage from hash table nodes (\crefrange{sec:ct}{sec:multithreading}).
\item \textbf{Implementation \& evaluation:} The Cuckoo Trie typically outperforms (or is comparable to) state-of-the-art indexes in both speed
and memory efficiency, across several datasets and workloads (\cref{sec:evaluation}). The Cuckoo Trie code is available at \url{https://github.com/cuckoo-trie/cuckoo-trie-code}.
\end{itemize}

\section{Background and terminology} \label{sec:background}

\noindentparagraph{Indexes.}
An \emph{ordered index} (henceforth, index) is a dictionary structure that supports \emph{point} operations (lookup,
insert, and delete) on key-value pairs as well as predecessor/successor queries, which can be used to perform
\emph{range} operations such as iteration over stored items by key order. Without loss of generality, we assume that
values are data records that also contain the key itself
(e.g., a database row)~\cite{hot}.%
\footnote{Otherwise, the index can allocate a record to hold both key and value~\cite{wormhole}.}

Indexes can be broadly classified into \emph{comparison-based}, \emph{tries}, and \emph{hybrids}~\cite{masstree,wormhole,hydra-list} of these classes.
Comparison-based indexes, such as B-tree~\cite{stx,fast,csb,k-ary,fast} and skip list~\cite{skiplist} variants, perform key
comparisons at each node. Their search time is typically $O(\log_B N$), where $B$ is the structure's branching factor and $N$ is the number of stored keys. Tries~\cite{art,hat-trie,hot,kiss-tree}
have $O(L)$ search time, where $L$ is the key length (in units of symbols, described below). Tries are thus attractive for data volumes where $\log_B N$ is larger than $L$.

\noindentparagraph{Tries.} \label{sec:tries}
A \emph{trie}~\cite{trie} is a tree-like data structure storing a sorted set of string keys. Each key is a sequence of \emph{symbols} over an alphabet $\Sigma$. For a key $k = k_0 ... k_{n-1}$ we denote by $\#k = n$ the length of $k$ and by $k[:i] = k_0 ... k_{i-1}$ the i'th prefix of $k$.

Each trie leaf corresponds to a key and each internal node to a key prefix. The \emph{name} of a node is the concatenation of all edge labels on the path to that node, and is a prefix of all of its descendants. A common optimization is \emph{path compression} \cite{patricia}: replacing a sequence of nodes having a single child each with a single edge having a multi-symbol label.

\noindentparagraph{Bucketized Cuckoo hashing.} \label{sec:bcht}
Cuckoo hashing \cite{cuckoo} is a hash table design in which each key can be stored in one of two possible table cells. To insert a key $k$, it is hashed with two hash functions, $h_1$ and $h_2$. If one of the cells $h_1(k)$ or $h_2(k)$ is empty, the key is placed there. If both are full, one of the keys in these cells is relocated to its other possible location. If this location contains a key, that key, too, is relocated, and so on. The main advantage of Cuckoo hash tables is their fast, constant-time lookups. Looking up a key $k$ requires checking just two independent reads of cells $h_1(k)$ and $h_2(k)$. However, Cuckoo hash tables can only support load factors of up to 50\% before insertions start to fail \cite{cuckoo}.

Bucketized Cuckoo hash table (BCHT) \cite{bucketized-cuckoo} is a development of the Cuckoo hash table that has better memory efficiency. A BCHT is built from constant-size \emph{buckets}, each containing up to $s$ items. As in Cuckoo hashing, each key $k$ is mapped to two buckets and is stored in one of them.
When inserting a key, if its two buckets are full then an item from one of the buckets is relocated to its other possible bucket,
which we term its \emph{alternate bucket}.

Bucketing greatly improves the load factor over regular Cuckoo hashing, while retaining its fast lookups. Even for 4-item buckets, load factors of over 90\% are achievable \cite{bucketized-cuckoo}.

\section{The need for MLP in index design} \label{sec:need-mlp}

Index operations are \emph{memory-bound}---that is, their execution time is dominated by memory access latency~\cite{mlp-def,masstree,art}.
Here, we show that much of this latency is \emph{potentially superfluous} and results from designs that do not leverage the ability of  modern hardware to exploit memory-level parallelism.

\subsection{Memory-level parallelism} \label{sec:mlp}

Modern processors can execute multiple instructions in parallel by means of \emph{out-of-order} execution~\cite{tomasulo1967efficient,hennessy2011computer}.
The processor computes dependencies between instructions as they are fetched from memory,
and an instruction can execute---regardless of its position in the instruction stream---as soon
as the instructions that it depends on have executed and produced their results.

Out-of-order execution can hide the latency of cache misses. This is important, as reading data from the last-level cache (LLC) or DRAM can take dozens to hundreds of clock cycles, respectively. While a memory read waits for its data to arrive, out-of-order execution allows subsequent memory reads whose operands do not depend on that read's result to execute in parallel with it.

The \emph{memory-level parallelism} (MLP) of a program is the average number of outstanding cache misses when there is at least one outstanding miss \cite{mlp-def}. Higher MLP is better, as it means that more of the memory access latency is overlapped.

\subsection{Index bottlenecks are caused by lack of MLP} \label{sec:no-index-mlp}

Virtually all state-of-the-art index designs share a common structure.
The index is an explicit hierarchical graph in memory, such that traversing through the graph gradually ``shrinks'' the reachable
key space until finally the target key is located or proven to be absent.
Such a hierarchical structure is fundamental for supporting ordered iteration/range queries, and underlies indexes
from skip lists~\cite{skiplist} through comparison-based search trees~\cite{fast,csb,masstree} to tries~\cite{art,hot}.

Unfortunately, this structure inherently lacks MLP.
The steps of traversing a search path cannot be overlapped, because the address of the next node on the path is read from the current node. Consequently, the amount of time the processor spends stalled on DRAM access is linear in the length of the search path and dominates index operation latency.

\begin{figure}
	\includegraphics[width=\linewidth]{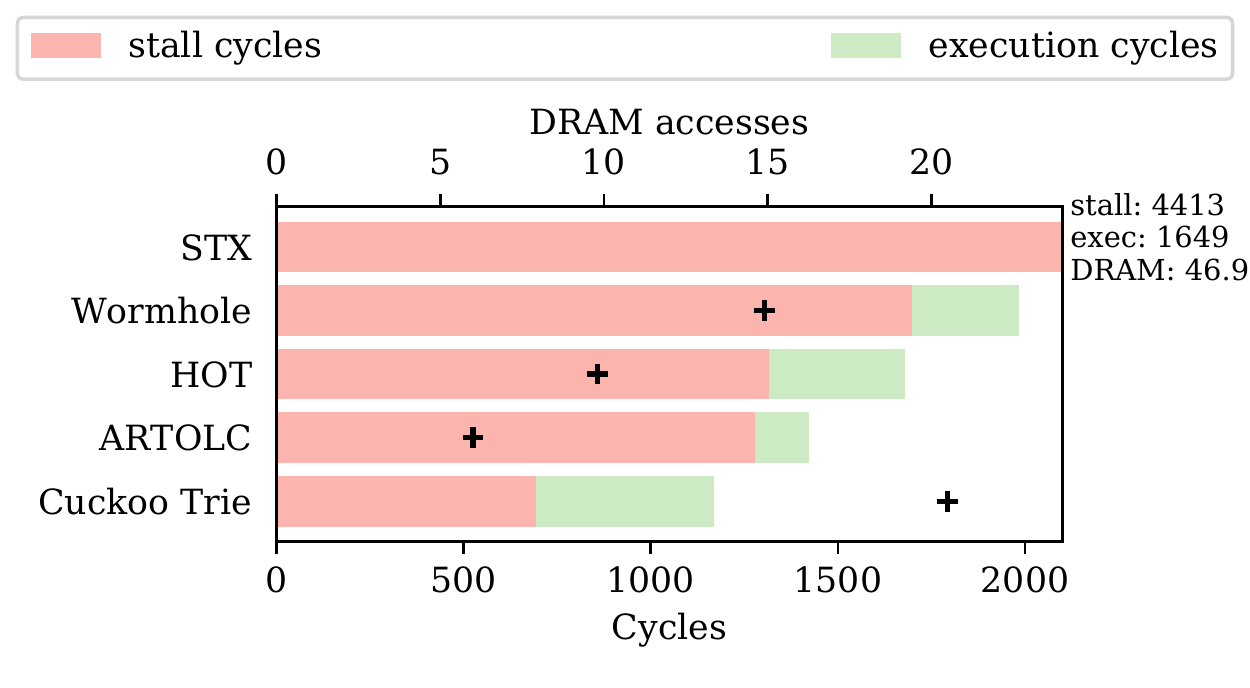}
    \vspace{-25pt}
	\caption{Cycles (bars) and DRAM accesses (crosses) per lookup for different indexes on a 200\,M random 8-byte keys dataset.}
	\label{fig:breakdown}
\end{figure}

\Cref{fig:breakdown} demonstrates the problem. We measure the latency (in clock cycles) of a key lookup in an index of
200\,M random 8-byte keys, for various state-of-the-art indexes and the Cuckoo Trie, on an Intel Skylake core.%
\footnote{The experimental setup and indexes used are described in detail in~\cref{sec:eval-setup}.}
Using the CPU performance counters, we break down the cycles into \emph{execution cycles}, where one or more instructions are executed, and \emph{stall cycles}, where the core just waits for DRAM accesses to complete.%
\footnote{Another cause of stall cycles is branch mispredictions. However, branch mispredictions account for less than 10\% of the stall cycles in all experiments.}

The Cuckoo Trie's MLP results in its random DRAM reads having effective latency ($\text{stall cycles} / \text{accesses}$) of 33.5\,cycles, about $3\times$ faster than in the indexes with serial accesses. Effective DRAM latency of the serial indexes varies, because some of their \emph{per-node} DRAM accesses may be overlapped by the CPU (e.g., reading a large node that spans multiple cache lines). Nevertheless, their lookup latency is  bounded from below by the latency of the serial path traversal. As a result, an entire Cuckoo Trie lookup takes less time than the other index lookups \emph{spend stalled on DRAM}---despite the Cuckoo Trie performing more DRAM accesses.

The takeaway is that without exploiting MLP, index designs face a performance barrier.
Much work focuses on accelerating the computational part of the lookup process, e.g., using SIMD instructions at each node~\cite{hot,k-ary,fast,parallel-skiplist} or efficient hash functions~\cite{wormhole}. But such optimizations can only save execution cycles, which are already a minority.

\section{The Cuckoo Trie} \label{sec:ct}

The Cuckoo Trie is a high-MLP, memory-efficient ordered index.
Conceptually, the Cuckoo Trie is a path-compressed trie (\cref{sec:tries}).
For space efficiency, it stores \emph{unique key prefixes}.
For each key $k$, the trie stores the shortest prefix $k[:i]$ such that $k[:i] \neq k'[:i]$ for any other key $k'$ in the trie;
the leaf associated with the unique prefix stores a pointer to the full key (more generally, to the record associated
with the key).

The Cuckoo Trie achieves high MLP by replacing the standard pointer-based trie with an implicit \emph{hashed representation}
that stores trie nodes in a Cuckoo hash table. This representation enables reading nodes along a path in parallel.
To achieve memory efficiency, the Cuckoo Trie uses a novel hash table design that \emph{eliminates} key storage (direct
or indirect) from hash table nodes. Here, we describe the single-threaded Cuckoo Trie; \cref{sec:multithreading} adds concurrency support.

\subsection{Obtaining MLP via a hashed trie representation} \label{sec:hashed-repr}

\begin{figure*}
	\includegraphics[width=\linewidth]{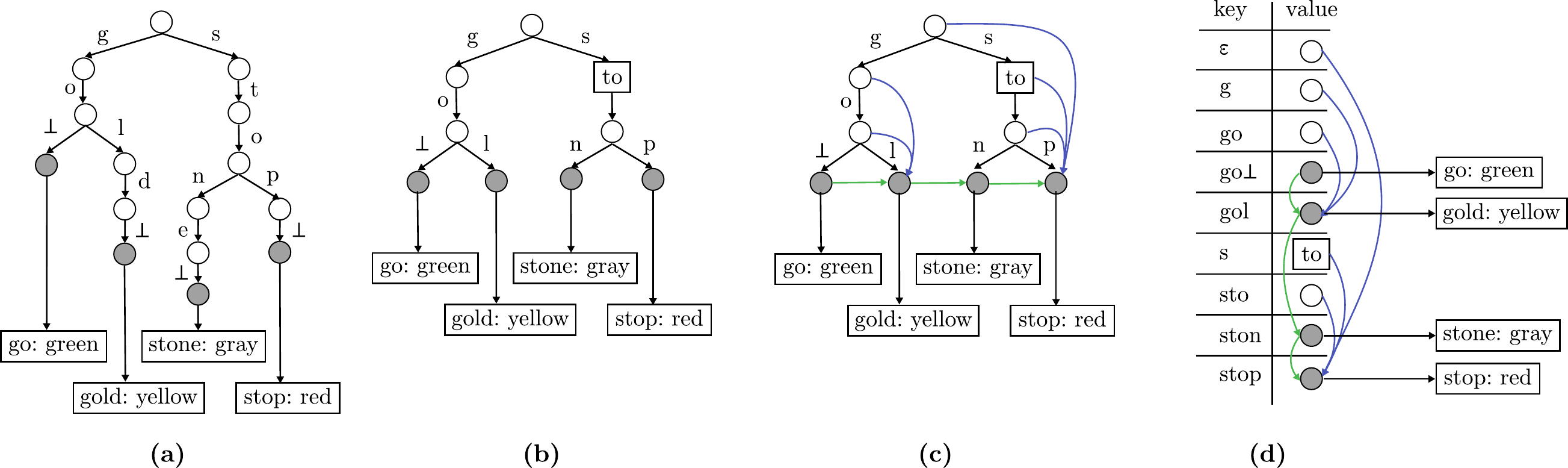}
	\caption{Cuckoo Trie vs. a regular trie. Starting from a regular trie containing 4 key-value pairs (a), a path-compressed trie of unique key prefixes is created by merging chains of nodes into ``jump nodes'' (marked with $\square$) and truncating subtrees containing a single leaf (b). Then \emph{next-leaf} (green) and \emph{subtree-max} (blue) locators (conceptually equivalent to pointers) are added (c). Finally, the hashed representation (d) removes the child pointers by storing the nodes in a hash table.}
	\label{fig:trie_types}
\end{figure*}

The Cuckoo Trie uses an implicit, pointer-free trie structure instead of an explicit pointer-based trie.
The idea is to store the trie nodes in a bucketized Cuckoo hash table (\cref{sec:bcht}).
The hash table stores key-value entries where the value is a trie node and the key is that node's name. We call this table the \emph{hashed representation} of the trie.
\Cref{fig:trie_types} depicts an example of the Cuckoo Trie hashed representation.

The hashed representation breaks the serial dependency of accessing nodes along a trie search path.
When searching for a key $k$, the names of the nodes that will be visited are known beforehand: these are the prefixes of $k$.
There are therefore no memory access dependencies between the hash table lookups of these nodes.
Consequently, waiting for a node $k[:i]$ to be read is no longer necessary to read its child $k[:i+1]$.
Moreover, due to the use of a BCHT, there are no dependencies within each hash table lookup.
Overall, the traversal has high MLP.

\noindentparagraph{Realizing MLP potential.}
Achievable MLP is limited by availability of hardware resources for instruction execution:

\noindent
\circled{1} The number of consecutive instructions that can execute out of order at any point in time is limited by the size of a hardware structure called a reorder buffer~\cite{johnson1991superscalar}, which is responsible for committing instruction results to architectural state in order.
Thus, if the number of instructions between two independent memory reads is larger than the reorder buffer, the processor will not be able to overlap the reads.

\noindent
\circled{2} The number of memory reads ``in flight'' is limited. Each in-progress memory access (except ones served by the L1/L2 cache) is tracked by a hardware structure called a miss status holding register (MSHR)~\cite{mshr1,mshr2}. The number of MSHRs in the processor thus limits MLP.

\noindentparagraph{Cuckoo Trie search.}
The Cuckoo Trie addresses the above challenges by using prefetch instructions. A search for a key begins by prefetching the hash table buckets
required for the first $D$ nodes of the path. (These instructions execute in parallel and commit before the data arrives, freeing reorder buffer space.) Then, each time the search moves one symbol down the path, the buckets for an additional node are prefetched. Therefore, at any instant, $D$ hash table nodes are being fetched from DRAM. We term $D$ the \emph{prefetch depth}.

As each memory access brings a whole cache line into the L1 cache, we design the hash table buckets to fit in a single 64-byte cache line (\cref{sec:hash-table}). This way, searching a node requires just two memory accesses, one for each bucket. We thus choose the prefetch depth such that $2 D$ is the number of parallel reads supported (i.e. the number of MSHRs).

\Cref{alg:lookup} shows the trie search algorithm.
In the $i$-th step, the \textsc{FindChild} function searches the hash table buckets for the node $k[:i]$. (The hash table search algorithm is detailed
in~\cref{sec:hash-table}; the $depth$ argument is related to path compression and can be ignored for now.) The trie search ends when
\textsc{FindChild} does not find the desired node, meaning that the target key is not in the trie, or when reaching a leaf.

\begin{algorithm}[b]
\footnotesize
\caption{Cuckoo Trie search path traversal}
\label{alg:lookup}
\begin{algorithmic}
\Statex \textbf{Parameters:}
\Statex $H$ - the hash function
\Statex $D$ - the prefetch depth
\Statex
\Function{Search}{$k$}
	\State restart:
	\For{$i \leftarrow 1$ \textbf{to} $D$}
		\State \textbf{prefetch} \Call{Buckets}{$H(k[:i])$} \Comment two BCHT buckets of $k[:i]$
	\EndFor
	\State $node \leftarrow Root$
	\State $depth \leftarrow 0$
	\For{$i \leftarrow 0$ \textbf{to} $\#k - 1$}
		\State \textbf{prefetch} \Call{Buckets}{$H(k[:D+i+1])$}
		\State $node', depth' \leftarrow$ \Call{FindChild}{$node,depth,k,i$}
		\If{$node' = $ \textbf{fail}}
			\State \textbf{goto} restart \Comment multithreaded conflict
		\EndIf
		\If{$node' = $ \textbf{null}}
			\State \textbf{return} $node$ \Comment reached a leaf
		\EndIf
		\State $node, depth \leftarrow node', depth'$
	\EndFor
	\State \textbf{return} $node$
\EndFunction
\end{algorithmic}
\end{algorithm}

\subsection{Hash table with key elimination} \label{sec:hash-table}

A key aspect of the Cuckoo Trie is its memory efficient hash table. Generally, any hashed representation obviates the need for child pointers in trie nodes: to find the child $c$ of a node named $s$ we simply search the key $s \cdot c$ in the table. While this makes the trie nodes smaller, the memory consumption of a hashed representation can still be significant.

Each hash table entry is a key-value pair, and while the values (trie nodes) are small, the keys (node names) can be large. Standard Cuckoo hash tables store the full key of each entry, either directly in the entry or indirectly, with the entry pointing to the key. Direct key storage is not memory efficient: each entry must have space for an entire key, despite the fact that most entries will hold only prefixes. Direct storage is also not compatible with variable-sized keys.%
\footnote{Indeed, current implementations using the hashed representation~\cite{mlp-index} only support short fixed-size keys.}
Indirect key storage hurts MLP, as reading the key incurs an extra DRAM access that depends on the hash table entry access.

\noindentparagraph{Solution: Key elimination.}
We support variable-sized keys using a novel bucketized Cuckoo hash table design. We use the fact that node names are prefixes of each other to avoid storing them in full. Instead, each entry contains just the last symbol of its prefix and a small amount of metadata. This way, the size of hash table entries is constant, regardless of the key size. We now show how this is done.

In Cuckoo hash tables, the key stored in each entry has two uses.
First, when relocating an entry, the key is required to \emph{compute the alternate bucket}.
Second, when searching for a key, if the key hash points to a non-empty table entry we have to \emph{verify the entry} by comparing the key stored in the entry with the target key. Otherwise, we might retrieve a key that is different from the target key but has the same hash value.
Our hash table handles the two uses separately.

\noindentparagraph{Relocations.}
To allow the computation of the alternate bucket without access to the full key, we instead store in each entry a small integer called a \emph{tag} from which the alternate bucket can be computed. We define the two buckets for a key $k$ in the following way: First, $k$ is hashed to a number $h(k) \in [0, S \cdot t)$ where $S$ is the number of buckets in the table and $t$ is a parameter controlling the tag size (in the Cuckoo Trie, $t = 16$). We split $h(k)$ into $B(k) = \floor{h(k) / t}$ and $T(k) = h(k)\; mod \; t$. Then the buckets are computed as
\begin{align*}
&B_1(k) = B(k) \\
&B_2(k) = B(k) + f(T(k))\; mod \; S
\end{align*}
with $f : [0,t) \rightarrow [0,S)$ a random function. The tag stored in each entry consists of $T(k)$ and a bit $b$ saying whether this entry is currently in its \emph{primary} ($B_1(k)$) or \emph{secondary} ($B_2(k)$) bucket. The tag, together with the entry's current bucket, uniquely determines $h(k)$ and in turn, the alternate bucket.

\noindentparagraph{Verification of entries.}
We use the fact that when searching a trie we always read the parent of a node before reading the node itself. When searching the hash table for a node named $p$ we can use the fact that the parent node, named $p[:\#p-1]$, was already found. Therefore, to verify that the key $k$ in a slot of bucket $B_i(p)$ is indeed $p$ we only have to check that:
\begin{enumerate}
	\item The last symbol of $k$ and $p$ is the same.
	\item The parent node of the node $k$ is the node $p[:\#p-1]$ that we already found.
\end{enumerate}

We can perform these checks using very little storage in each entry by using a special kind of hash function, which we term a
\emph{peelable} hash function.
Intuitively, being a peelable hash function means that given the hash of a string $x$ and its last symbol we can ``peel'' the last symbol and compute the hash of $x[:\#x-1]$. Formally:
\begin{unnumbered_definition}
A function $h : \Sigma^* \rightarrow \mathbb{N}$ is called \emph{peelable} if there is a function $P_h : \mathbb{N} \times \Sigma \rightarrow \mathbb{N}$ such that ${P_h(h(x \cdot c), c) = h(x)}$ for all $x \in \Sigma^{*}, c \in \Sigma$.
\end{unnumbered_definition}
The Cuckoo Trie uses the following peelable function:
\begin{align*}
&h(\epsilon) = 0 \\
&h(x \cdot c) = \floor{\frac{h(x) \oplus c}{R}} + \frac{S t}{R} ((h(x) \oplus c) \; mod \; R)
\end{align*}
where $x$ is a string, $c$ a symbol, and $R$ a small power of two (e.g. $R = 2^5$).  $S$, the number of buckets, is chosen such that $R$ and $|\Sigma|$ divide $S t$.

This hash function is peelable because we can set
$$P_h(y, c) = c \oplus (R \cdot (y \; mod \; \frac{St}{R}) + \floor{\frac{yR}{St}})$$
and then $P_h(h(x \cdot c), c) = h(x)$ as required.

Each entry contains three fields for entry verification:
\begin{itemize}
\item \emph{last\_symbol}: The last symbol of the key $k$ stored in that entry.
\item \emph{color}: A small integer chosen on key insertion such that it is unique among entries with the same key hash
(i.e., $2B$ colors are needed for $B$-entry buckets).
\item \emph{parent\_color}: The color of the entry that contains the key $k[:\#k-1]$.
\end{itemize}

Given an entry to verify, checking (1) is trivial, by looking at the \emph{last\_symbol} value. To check (2), we only have to check that the tag is correct and that the \emph{parent\_color} field matches the color of the parent entry. The tag, together with the entry location, determines the key hash for the entry. This, together with \emph{last\_symbol} determines the parent entry key hash, using the fact that $h$ is peelable.%
\footnote{
Note that the verification relies on the existence of the peel function $P_h$ for its correctness, but does not compute it. Therefore, when choosing $h$ we do not have to take the efficiency of $P_h$ into account.}
Finally, the \emph{parent\_color} uniquely identifies the parent entry.%

\begin{figure}
	\includegraphics[width=\linewidth]{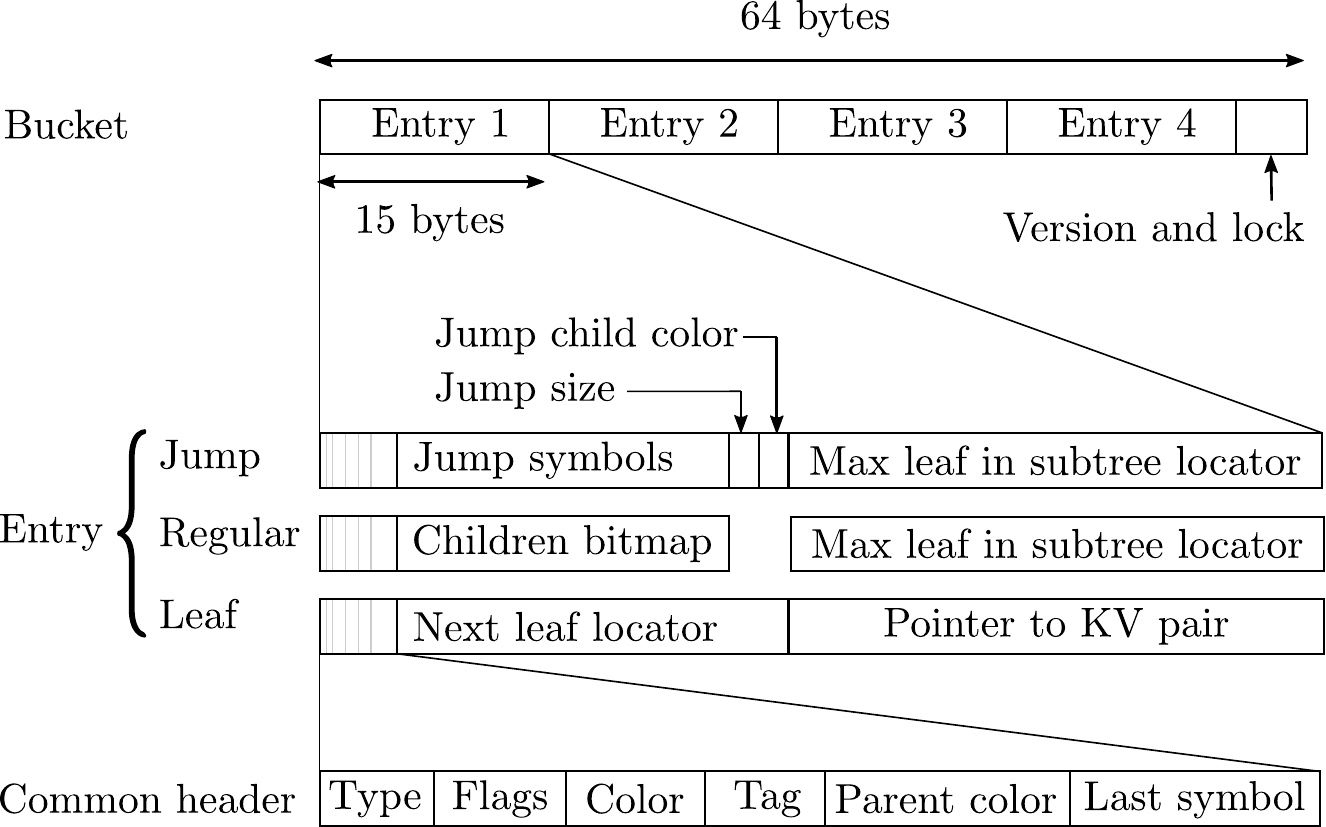}
	\caption{Layout of hash table entries and buckets. A locator of a node $n$ with name $x$ is $(h(x),n.color)$, which uniquely identifies $n$'s entry.}
	\label{fig:layout}
\end{figure}

\Cref{fig:layout} shows the structure of the hash table buckets and entries.
Each 64-byte bucket contains four 15-byte entries and multithreading-related metadata (\cref{sec:multithreading}).
Each entry specifies the type of trie node it stores and the aforementioned fields.
(``Jump nodes'' implement path compression and are described in~\cref{sec:path-compression}.)
\Cref{alg:hashtable} shows the \textsc{FindChild} algorithm. Given the current node $n$ with name $k[:i]$, \textsc{FindChild}
searches the hash table for the child $k[:i+1]$ by calling \textsc{SearchByParent}, which performs the above verification.
To stop a search for a non-existent child, we leverage the fact that each internal node contains a list of its children,
which is needed for range queries (\cref{sec:range-iters}) and multithreading (\cref{sec:multithreading}).
The list is stored as a bitmap with one bit per possible child (next symbol).

\begin{algorithm}
\footnotesize
\caption{Cuckoo hash with key elimination}
\label{alg:hashtable}
\begin{algorithmic}
\Statex \textbf{Parameters:}
\Statex $B$ - the table, containing $S$ buckets.
\Statex $T$ - the tag size.
\Statex $H:\Sigma^* \rightarrow [0,ST)$ - the hash function.
\Statex $M$ - a table of $T$ random numbers, $M[i] \in [0,ST)$.
\Statex
\Function{Buckets}{$h$}
	\State $b_1 \leftarrow h / T$
	\State $b_2 \leftarrow (b_1 + M[h \; mod \; T]) \; mod \; S$
	\State \textbf{return} $b_1, b_2$
\EndFunction
\Statex
\Function{EntriesFor}{$h$}
	\State $b_1, b_2 \leftarrow \Call{Buckets}{h}$
	\State $tag \leftarrow h \; mod \; T$
	\State $S_1 \leftarrow \{n \in B[b_1] \; | \; n.tag = tag \wedge n.is\_primary\}$
	\State $S_2 \leftarrow \{n \in B[b_2] \; | \; n.tag = tag \wedge \neg n.is\_primary\}$
	\State \textbf{return} $S_1 \cup S_2$
\EndFunction
\Statex
\Function{SearchByParent}{$hash, last\_sym, color$}
	\For{$c$ \textbf{in} \Call{EntriesFor}{$hash$}}
		\If{$c.last\_symbol = last\_sym$}
			\If {$c.parent\_color = color$}
				\State \textbf{return} $c$
			\EndIf
		\EndIf
	\EndFor
	\State \textbf{return null}
\EndFunction
\Statex
\Function{SearchByColor}{$hash, last\_sym, color$}
	\For{$c$ \textbf{in} \Call{EntriesFor}{$hash$}}
		\If{$c.last\_symbol = last\_sym$}
			\If {$c.color = color$}
				\State \textbf{return} $c$
			\EndIf
		\EndIf
	\EndFor
	\State \textbf{return null}
\EndFunction
\Statex
\Function{FindChild}{$n, d, k, i$}
	\If{$n$ \textbf{is} leaf}
		\State \textbf{return null}, 0
	\EndIf
    \Repeat
	   \State $v \leftarrow bucketOf(n).version$
    \Until $v\text{ is even}$
	\State $h \leftarrow H(k[:i+1])$ \Comment Hash value of the child
	\State restart:
	\If{$n$ \textbf{is} regular node}
		\If{$n.bitmap[k_i] = 0$}
			\State \textbf{return null}, 0
		\EndIf
		\State $c \leftarrow$ \Call{SearchByParent}{$h, k_i, n.color$}
	\Else \Comment $n$ is a jump node
		\If{$n.symbols[d] \neq k_i$}
			\State \textbf{return null}, 0
		\EndIf
		\If{$d + 1 < n.size$}
			\State \textbf{return} $n, d + 1$
		\EndIf
		\State $c \leftarrow$ \Call{SearchByColor}{$h, k_i, n.child\_color$}
	\EndIf
	\If{$v \neq bucketOf(n).version$}
		\State \textbf{return fail}, 0
	\EndIf
	\If{$c =$ \textbf{null}}
		\State \textbf{goto} restart
	\EndIf
	\State \textbf{return} $c$, 0
\EndFunction
\end{algorithmic}
\end{algorithm}

\subsection{Path compression} \label{sec:path-compression}

In the Cuckoo Trie, chains of nodes each having only a single child are compressed into \emph{jump nodes} (\cref{fig:trie_types}b). A jump node contains the labels of all compressed edges. To keep the maximal jump node size fixed, long chains are split and each part is compressed to a separate jump node.

In~\cref{alg:lookup}, when a search iteration lands at a jump node, the $depth$ variable is initialized to zero and subsequently used to compare the searched keys' symbols to the jump node's label, symbol by symbol, until there is either a mismatch (key not found) or the traversal moves to the jump node's child, which is located using \textsc{SearchByColor} (\cref{alg:hashtable}).
As a jump node contains multiple symbols, we cannot ``peel'' its child's hash value to obtain the jump node's hash, and so cannot find the child by its \emph{parent\_color} field as for regular nodes. Instead, a jump node contains the color of its sole child, thereby identifying it uniquely.

\subsection{Lookups and range iterations} \label{sec:range-iters}

\noindentparagraph{Lookups.}
Lookups perform a trie search (\cref{alg:lookup})---as opposed to simply looking up the key in the hash table---because the Cuckoo Trie stores unique key prefixes and not full keys. (This is a design choice aimed at balancing index speed and memory consumption; see~\cref{sec:justify-unique}.) If the trie search stops at a leaf, the key the leaf points to is read and compared to the target key. If there is a match, the corresponding value is returned. Otherwise, the lookup fails.

\noindentparagraph{Range iteration.}
Executing a range query asking for the keys between $k_1$ and $k_2$ requires two primitives from an ordered index. First, we need to find the range start key, which is the smallest key in the index larger than $k_1$. Second, once this key is found, we need an efficient way to repeatedly move to the next key until reaching $k_2$ or a larger key.

\begin{figure}
	\includegraphics[width=0.7\linewidth]{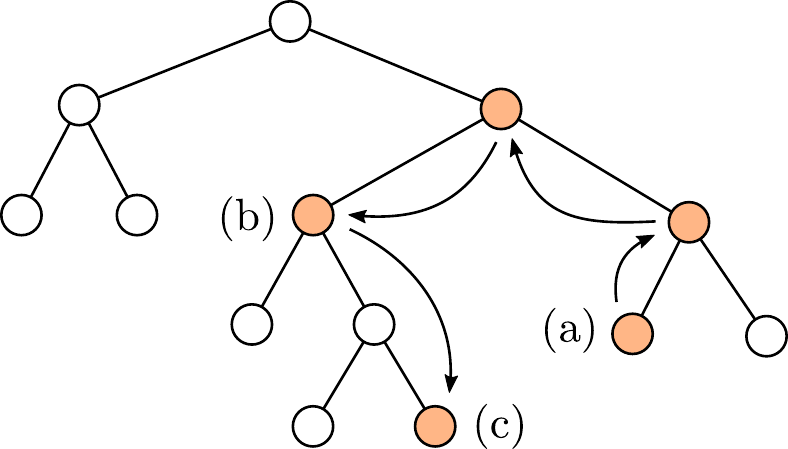}
	\caption{Predecessor search. To find the predecessor of a leaf (a) the trie is traversed upwards until a node with a smaller child is found. The \emph{subtree-max} locator from this child (b) is followed to reach the predecessor (c).}
	\label{fig:predecessor}
    \vspace{-10pt}
\end{figure}

\vspace{0.5\baselineskip}\noindent
\emph{Range start:} Finding the start leaf is done using the standard predecessor search algorithm for tries. We search for $k_1$ in the trie and if not found, go to the predecessor of the deepest node reached during the search. This entails ascending up the path until finding a node with a left sibling (checked via its parent's bitmap list of children), and then finding the maximal leaf in the sibling's subtree. To accelerate the search, each internal node contains a conceptual ``pointer,'' termed a \emph{locator}, to the maximal leaf in its subtree. This ``pointer'' allows skipping directly to the maximal leaf, which would otherwise require multiple memory accesses (one for each level) that cannot be parallelized, as the key of the maximal leaf is not known. \Cref{fig:predecessor} depicts the predecessor search.

As entries in the BCHT can be relocated at any time, a locator does not contain the memory address of the node it points to. Instead, it contains its key hash and color, which allow finding it using only two bucket reads even after relocations.

\vspace{0.5\baselineskip}\noindent
\emph{Moving to next key:}
In tries, moving from one key to the next in the range requires a costly up-and-down traversal. Therefore, we maintain the leaves in a linked list in sorted order. As with the \emph{subtree-max} pointers, linked list pointers use locators instead of a memory address.

\noindentparagraph{MLP in range iteration.}
Using a linked list of leaves for range queries requires one memory access for each range element, and each access is dependent on the previous one, slowing down range scans. The usual solution, employed in \cite{hot,wormhole}, is to truncate paths at some prefix, so that a trie leaf contains multiple keys. Fetching a leaf from DRAM thus enables scanning multiple keys, thereby amortizing that cost. However, this solution also complicates the insertion procedure, as to make the amortization pay off, keys have to be moved between leaves to keep each leaf well-populated.

To keep the insertion procedure simple, the Cuckoo Trie stores a single key in each leaf. To avoid slowing down range iteration, we use the fact that usually the system using the index performs some work for each key iterated over. Each time we return a key pointer to the system, we prefetch the key pointed by its node's \emph{next} pointer. This way, reading each key is overlapped with the system's work. The work done by a typical database usually takes longer than the memory latency. Therefore, as will be demonstrated in~\cref{sec:system-benchamrk}, the latency associated with following the \emph{next} pointer is completely hidden, and in realistic settings, traversal speed is comparable to that of the multiple-keys-per-leaf approach.

\subsection{Insertions and deletions} \label{sec:point-ops}

Inserting a new key begins by descending according to its symbols, similar to a lookup. Eventually, the descent will reach a final node from which it can no longer advance. There are three cases: If this node is a regular node, a new leaf is added under it which points to the new key. If the final node is a leaf, the key it points to is compared to the new key, and it is replaced with a chain of nodes representing their common prefix. Under the last node of this chain, two leaves are created: one for the original key and one for the new one. Finally, if the final node is a jump node, it is split into regular nodes and then treated as a regular node.

After the leaf for the new key is created, the pointers in the tree are updated. First, it is inserted into the linked list of the leaves (this requires finding its predecessor). Then, if the new leaf is maximal in one or more subtrees, the appropriate \emph{subtree-max} pointers are updated.

Deletion is similar to insertion: First, the leaves corresponding to the deleted key and its predecessor are found. Then, the leaf corresponding to the deleted key is marked absent in the bitmap in its parent and removed from the linked list. If the deletion results in a subtree that contains a single leaf (a ``tail''), the subtree is replaced by that leaf. Finally, \emph{subtree-max} pointers are updated and the deleted leaf is removed from the trie.

\subsection{Design rationale} \label{sec:justify-unique}

A key design goal of the Cuckoo Trie is to \emph{balance} index speed and memory efficiency. We thus purposefully
sacrifice speed optimization opportunities that trade off memory efficiency and vice versa. Here, we describe these decisions
and their implications. The Cuckoo Trie can, of course, be adapted to systems where speed is more important than memory efficiency
(or vice versa) by changing these trade-offs.

\noindentparagraph{Unique key prefixes.}
We can significantly improve point operation latency by storing full keys instead of unique prefixes in the Cuckoo Trie.
For instance, lookups could simply search the hash table for the full key's entry with two overlapping DRAM accesses (plus the subsequent
key comparison). However, this would dramatically increase index size. For example, on a dataset of 200\,M random 8-byte keys, the Cuckoo Trie uses an average of 1.25 nodes per key.%
\footnote{Each key requires one leaf, and, because the keys are random, internal nodes have many children, resulting in an average of 0.25 internal nodes per leaf (this was determined experimentally).}
Storing full keys would require (at least) one additional node per key, increasing the index size by 80\%.

\noindentparagraph{Symbol size.}
The Cuckoo Trie's symbol size is also a trade-off. The latency of a search depends on the symbol size.
Due to hardware limitations, we can only fetch $D$ levels in parallel (\cref{sec:hashed-repr}).
So the search essentially requires $L/D$ DRAM round trips to complete, where $L$ is the total number of levels a search
reads---and $L$ decreases as the symbol size increases. On the other hand, increasing the symbol size increases
the size of the internal nodes' child bitmap, causing fewer entries to fit in a single-cache line bucket, and
ultimately to bucketing being ineffective.

\subsection{Limitations} \label{sec:limitations}

Compared to conventional index designs, the Cuckoo Trie's focus on MLP leads to certain limitations, discussed below.

\noindentparagraph{Path compression only saves space, not DRAM accesses.}
In conventional tries, reading one compressed (``jump'') node allows traversing multiple symbols and thereby reduces the number of nodes fetched from DRAM.
The Cuckoo Trie, in contrast, attempts to read all prefixes of a key up front---even ones that turn out to share a jump node with earlier prefixes. Therefore, traversing a part of a key compressed into a jump node issues the same number of DRAM accesses (prefetches) as when each symbol is represented by its own node.
The Cuckoo Trie's jump nodes thus serve only to reduce memory usage.

\noindentparagraph{Keys with long common prefixes.}
The Cuckoo Trie is less competitive on such data sets.
Because path compression does not reduce the number of DRAM accesses, when keys have long common prefixes, the number of non-overlapped DRAM round trips required to perform a trie search can exceed the
number of serial DRAM accesses performed by indexes whose path compression saves DRAM accesses (\cref{sec:eval:point}).

\noindentparagraph{Superfluous memory accesses.}
Making the DRAM accesses independent of each other can result in three types of superfluous DRAM accesses, which fetch cache lines that
end up not being needed by the search:
\circled{1} Once the leaf corresponding to the searched key's unique prefix is located, there may be prefetches in flight for subsequent symbols of the key.
\circled{2} For each trie node traversed, both buckets in which it might reside are fetched. For nodes located in their primary bucket, fetching the alternate bucket is superfluous (this limitation is inherited from Cuckoo hashing).
\circled{3} Fetches of buckets associated with symbols that are compressed into a jump node are superfluous. Importantly, however, once the traversal leaves the jump node, it again benefits from the prefetching of nodes associated with subsequent, uncompressed symbols.

Fortunately, modern hardware is designed to absorb the high DRAM access rates required by CPUs to leverage MLP. As a result, we find that the Cuckoo Trie's overall memory bandwidth needs are far from hardware limits (\cref{sec:eval:membw}).

\section{Concurrency support} \label{sec:multithreading}

The Cuckoo Trie provides linearizable~\cite{linearizability} insert, delete, lookup, and predecessor/successor search operations.%
\footnote{Linearizability proofs appear in the full version of this paper~\cite{adar-thesis}.}
As in other state-of-the-art indexes~\cite{masstree,hot}, range iteration is not atomic with respect to insertions/deletions,
because it is performed by a sequence of successor searches from the range start key.

The Cuckoo Trie uses standard optimistic lock-based concurrency control techniques~\cite{snaptree,masstree,rowex} for multicore scalability.
Searches are read-only and do not acquire locks. Writing operations lock only affected buckets (located with a
read-only search). Searches rely on per-bucket version numbers, which writers modify, to verify they have observed a consistent state of the trie, and restart the operation otherwise.

We use per-buckets versions and locks, co-located in the same 32-bit word (\cref{fig:layout}), for space efficiency. A bucket's version is incremented twice for each update, before and after the update, forming a seqlock~\cite{seqlock}. This allows a search to atomically read the 15-byte hash table entries, by checking the version number before and after the read, and retrying if the number is odd or if it changed between the two checks. (This happens in \textsc{EntriesFor} in \cref{alg:hashtable}, but we omit the pseudo code.)

\noindentparagraph{Updates.}
An insert/delete operation locates all nodes it will write to using the thread-safe search and predecessor search described below.
It then simultaneously locks the buckets and verifies they have not changed in the meantime by using an atomic \texttt{compare-and-swap}
on the buckets' version/lock word. An insertion (resp., deletion) locks the parent and predecessor of the inserted (resp., deleted) node and any ancestors whose \emph{subtree-max} pointer is affected. A deletion also locks the bucket of the deleted leaf.

After acquiring all the locks, the nodes are written to and then all locks are released.
New leaves are created locked and only released when the update is finished.
Nodes are updated in the order described in~\cref{sec:point-ops}. This order keeps the trie in a consistent state during most of the update's execution, allowing for concurrent searches. The only fields that can have an inconsistent value are the \emph{subtree-max} and \emph{next} pointers. This happens when a node is added or deleted but the pointers were not yet updated. To account for this inconsistency, writers mark a leaf with a ``dirty'' flag if it is pointed by a stale \emph{subtree-max} pointer or if it contains a stale \emph{next} pointer. If a search or scan reaches a dirty leaf, it restarts its operation.
After the pointers are made consistent, the ``dirty'' flag is cleared.

In addition, deletions use the ``dirty'' flag as a ``deleted'' marker, setting it on the deleted leaf after acquiring all locks. This simultaneously makes the leaf unreachable for lookups and range scans, and invalidates any \emph{subtree-max} pointers that point to it.

\noindentparagraph{Key searches.}
Given atomic entry reads, the only change required to the single-threaded search procedure is to account for concurrent relocation or deletion of the currently read node. This is done by repeatedly searching for the node in its two possible buckets and then checking the version number of its parent. If the version number changes, the search fails and the operation it was part of is restarted (see \textsc{FindChild} in~\cref{alg:hashtable}).
If there is a concurrent relocation, the node will eventually be found in a later iteration. If the node is deleted, the deleting thread will change the parent's version when it removes the deleted child from its bitmap, and the search will restart.

\noindentparagraph{Predecessor searches.}
Both updates and range scans perform predecessor searches. Predecessor search in the presence of concurrent modifications is implemented using validation: after finding a node's predecessor, the operation verifies that the version numbers of all buckets it traversed
(colored nodes in~\cref{fig:predecessor}) did not change. This ensures the consistency of the path when the predecessor was reached (or, for updates, when it gets locked).

\noindentparagraph{Range iteration.}
When following the \emph{next} pointer from a leaf with key $k_1$ to a leaf with key $k_2$, leaf $k_2$ might become an internal node or be deleted before we read it. This is detected by re-reading the version of $k_1$ and verifying that it did not change. If $k_1$ did change, the iterator is resynchronized by performing a new search for $k_1$, starting from the root, and following its updated \emph{next} pointer.

\noindentparagraph{Resizing.}
Resizing can be performed in a thread-safe way---without blocking searches from using the old hash table while the resize process copies it into a larger one---using an existing technique of marking a bucket before copying it to the new table, to prevent the bucket from being subsequently modified in the old table~\cite{resizing-hash}.

\section{Evaluation} \label{sec:evaluation}

We compare the Cuckoo Trie to state-of-the-art ordered indexes with respect to performance, memory efficiency, and scalability,
on various key distributions and workloads.

\subsection{Experimental setup} \label{sec:eval-setup}

\noindentparagraph{Platform.}
We use a server with two NUMA nodes. Each node contains an Intel Xeon Gold 6132 (Skylake) processor with 14 2.6\,GHz cores and 96\,GB DDR4-2666 memory. The NUMA nodes are connected with an Ultra Path interconnect (UPI). Memory allocations are interleaved between the nodes. Hyper-threading and Turbo-Boost are disabled.
In multithreaded runs, threads are split equally between the nodes. Code is compiled using GCC 7.5.0 and run on Ubuntu 18.04.
Reported numbers are averaged over 3 runs (all measurements are within $\pm 8\%$ of the average).

\noindentparagraph{Indexes.}
We compare the Cuckoo Trie to the following:
\begin{itemize}[leftmargin=*]
\item \textbf{HOT}~\cite{hot}: The Height Optimized Trie keeps the trie shallow and balanced by considering a variable number of key bits in each node.

\item \textbf{Wormhole}~\cite{wormhole}: A combined trie and optimized B-tree achieving $O(\log L)$ theoretical complexity for $L$-byte keys.

\item \textbf{ARTOLC}~\cite{building-bwtree}: A thread-safe implementation of the Adaptive Radix Tree~\cite{art} using optimistic lock coupling. Was found in~\cite{building-bwtree} to be faster than many other indexes.\footnote{We use the code from \url{https://github.com/wangziqi2016/index-microbench}.}

\item \textbf{STX}~\cite{stx}: An optimized in-memory B-tree. We use the improved implementation from the TLX library~\cite{tlx}.

\item \textbf{MlpIndex}~\cite{mlp-index}: An MLP-aware index using a basic hashed trie representation. It only supports 8-byte keys and single-threaded operation, and so is omitted from most runs.
\end{itemize}

We use the jemalloc~\cite{jemalloc} memory allocator and 2\,MB huge pages for all memory allocations.
All indexes store pointers to key-value pairs and have to dereference them to access the actual keys and values.
The Cuckoo Trie is configured with a  hash table load factor of 85\%, 5-bit symbols, and a prefetch depth $D=5$, which was experimentally found to give the best results on our system, whose cores
have 12 MSHRs each~\cite[Appendix A]{intel-mlp}. Our current implementation omits deletions and automatic resizing.

\noindentparagraph{Datasets.}
We use datasets of varying key lengths and unique prefix distributions (\cref{table:dataset}), including worst-case ones for the Cuckoo Trie. \emph{rand-8} and \emph{rand-16} consist of random 8- and 16-byte keys, respectively. The other datasets are based on real-world key distributions: \emph{osm} contains 64-bit encodings of random OpenStreetMap locations (osmc64~\cite{sosd}); \emph{az} keys are tuples of item ID, user ID, and time describing Amazon product reviews (Az1~\cite{wormhole}); and \emph{reddit} is a list of all usernames on Reddit as of 2018.%
\footnote{\url{https://files.pushshift.io/reddit/RA\_2018-09.gz}}
The \emph{az} dataset is a worst case for the Cuckoo Trie, as it contains keys with long common prefixes (\cref{sec:limitations}).
All datasets are shuffled randomly and have duplicate keys removed.
In all datasets, the key set and index are each at least 1\,GB in size and thus far exceed the system's aggregated LLC capacity.

\begin{table}
\centering
\small
\begin{tabular}{c|
				>{\centering\arraybackslash}m{1.6cm}
				>{\centering\arraybackslash}m{2.4cm}
				>{\centering\arraybackslash}m{2cm}}
\toprule
\textbf{Name}    & \textbf{Avg. key size (bytes)} & \textbf{Avg. unique prefix size (bits)} & \textbf{No. of keys} \\
\midrule
rand-8  & 8    & 28.9  & 200\,M \\
rand-16 & 16   & 28.9  & 200\,M \\
osm     & 8    & 36.8  & 200\,M \\
az      & 35.7 & 138.2 & 82\,M  \\
reddit  & 10.9 & 63.7  & 71\,M  \\
\bottomrule
\end{tabular}
\caption{Datasets used in the experiments.}
\label{table:dataset}
\vspace{-10pt}
\end{table}

\noindentparagraph{Workloads.}
Most experiments are based on standard workloads from the Yahoo! Cloud Serving Benchmark (YCSB)~\cite{ycsb}. Each workload specifies a mix of operation types intended to simulate a real-world use case (\cref{table:ycsb}). We also evaluate the LOAD workload, which loads the dataset into the index. Unless otherwise noted, we use the same parameters as the default YCSB implementation, and a uniform query distribution.
Zipfian distribution results appear in the full version of this paper~\cite{adar-thesis}.

\begin{table}
\centering
\small
\begin{tabular}{l|
				>{\arraybackslash}m{6cm}}
\toprule
\textbf{Workload}    & \textbf{Description} \\
\midrule
LOAD  & 100\% insert        \\
A  & 50\% lookups, 50\% updates of existing keys  \\
B  & 95\% lookups, 5\% updates of existing keys  \\
C  & 100\% lookups  \\
D  & 5\% inserts, 95\% lookups (recently-inserted key) \\
E  & 5\% inserts, 95\% scans (random size in $[1,100]$) \\
F  & 50\% lookups, 50\% read-modify-write operations \\
\bottomrule
\end{tabular}
\caption{YCSB workloads.}
\label{table:ycsb}
\vspace{-20pt}
\end{table}

\subsection{Point operation performance} \label{sec:eval:point}

\begin{figure}
\includegraphics[width=\linewidth]{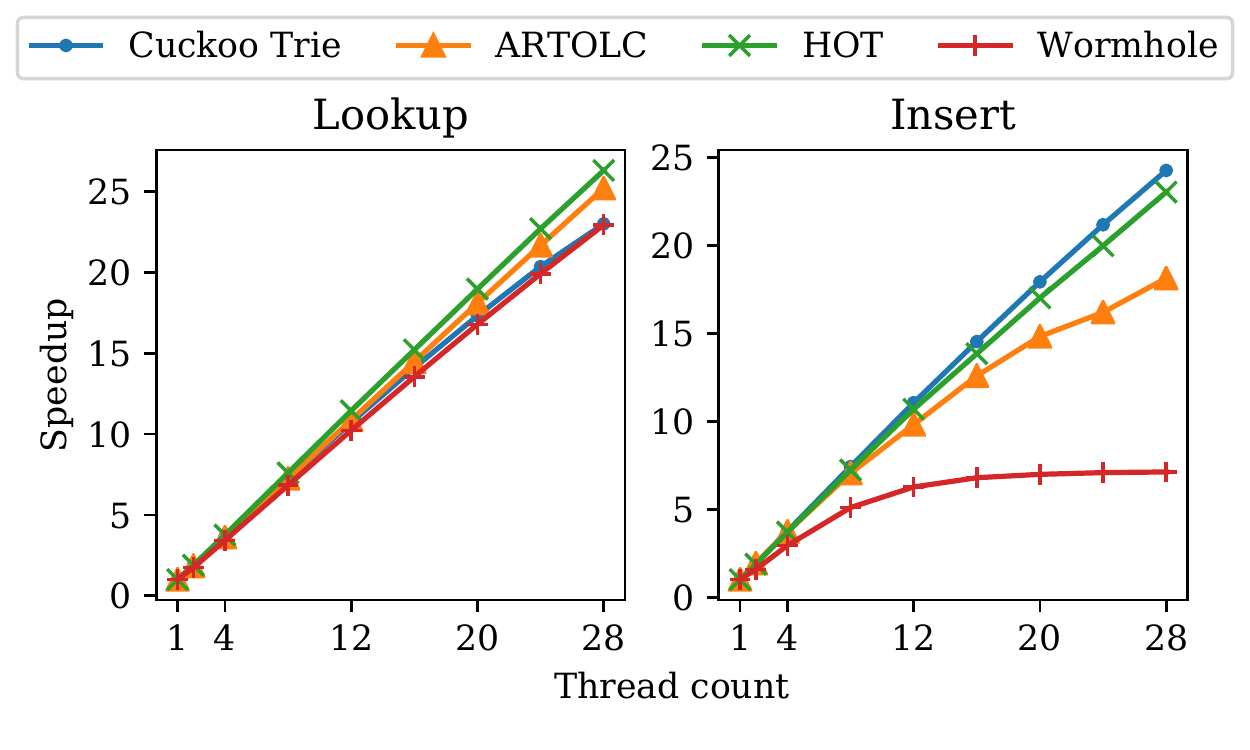}
\vspace{-30pt}
\caption{Insert \& lookup scalability on the \emph{rand-8} dataset.}
\label{fig:scalability}
\vspace{-10pt}
\end{figure}

\noindentparagraph{Multicore scalability.}
We measure lookup and insertion throughput of each dataset with a varying number of threads. To measure lookup performance, we use the YCSB-C (lookup only) workload, and set the workload size to 10\,M queries per thread. Insertion is measured using the LOAD workload (inserting the whole dataset). \Cref{fig:scalability} shows the speedup each index achieves with respect to its single-threaded throughput on the \emph{rand-8} dataset. STX is omitted as it only supports single-threaded operation. Results of other datasets are also omitted, as they are qualitatively the same as in \emph{rand-8}.

All indexes scale almost linearly in both experiments, except for Wormhole and ARTOLC, whose insertion throughput becomes saturated with many threads.

\begin{figure*}[t]
	\includegraphics[width=0.8\linewidth]{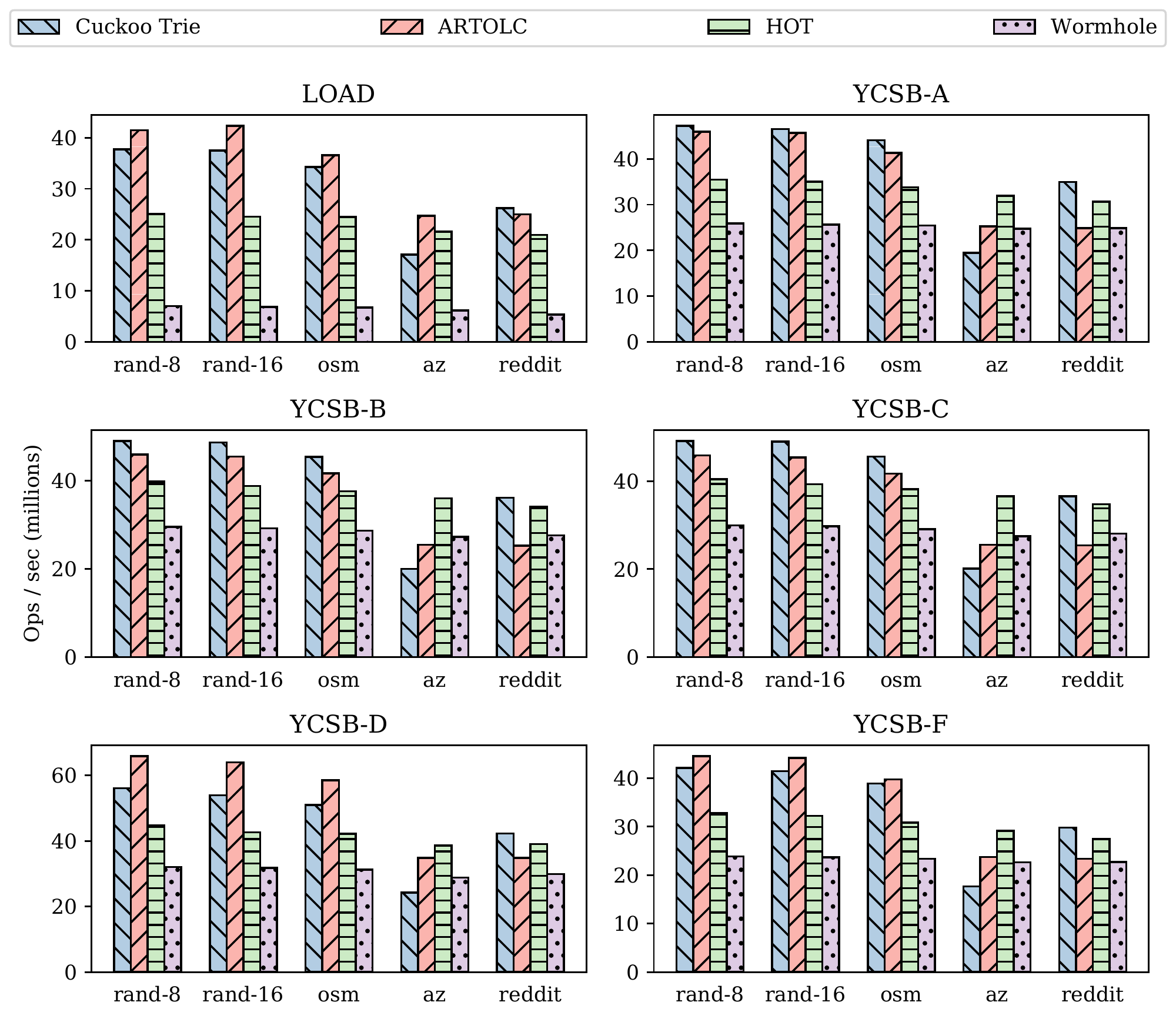}
    \vspace{-15pt}
	\caption{Multithreaded (28 cores) throughput of YCSB point-operation workloads.}
    \label{fig:ycsb_point_mt}
\end{figure*}

\noindentparagraph{Throughput.}
\Cref{fig:ycsb_point_1t} and \cref{fig:ycsb_point_mt} show the throughput measured on single- and multithreaded (using all cores) point operation workloads, respectively.
In the following, we focus on single-threaded execution, which does not penalize non-scalable indexes and,
for scalable indexes (including the Cuckoo Trie), its results extrapolate to the multithreaded execution.
On most workload-dataset combinations, the Cuckoo Trie outperforms the other indexes---by up to 35\% (HOT), 20\% (ARTOLC), 50\% (Wormhole) and 360\% (STX) on YCSB-C with \emph{osm}.
The cases where it is slower are due to one of three causes:

\begin{figure*}
\includegraphics[width=0.8\linewidth]{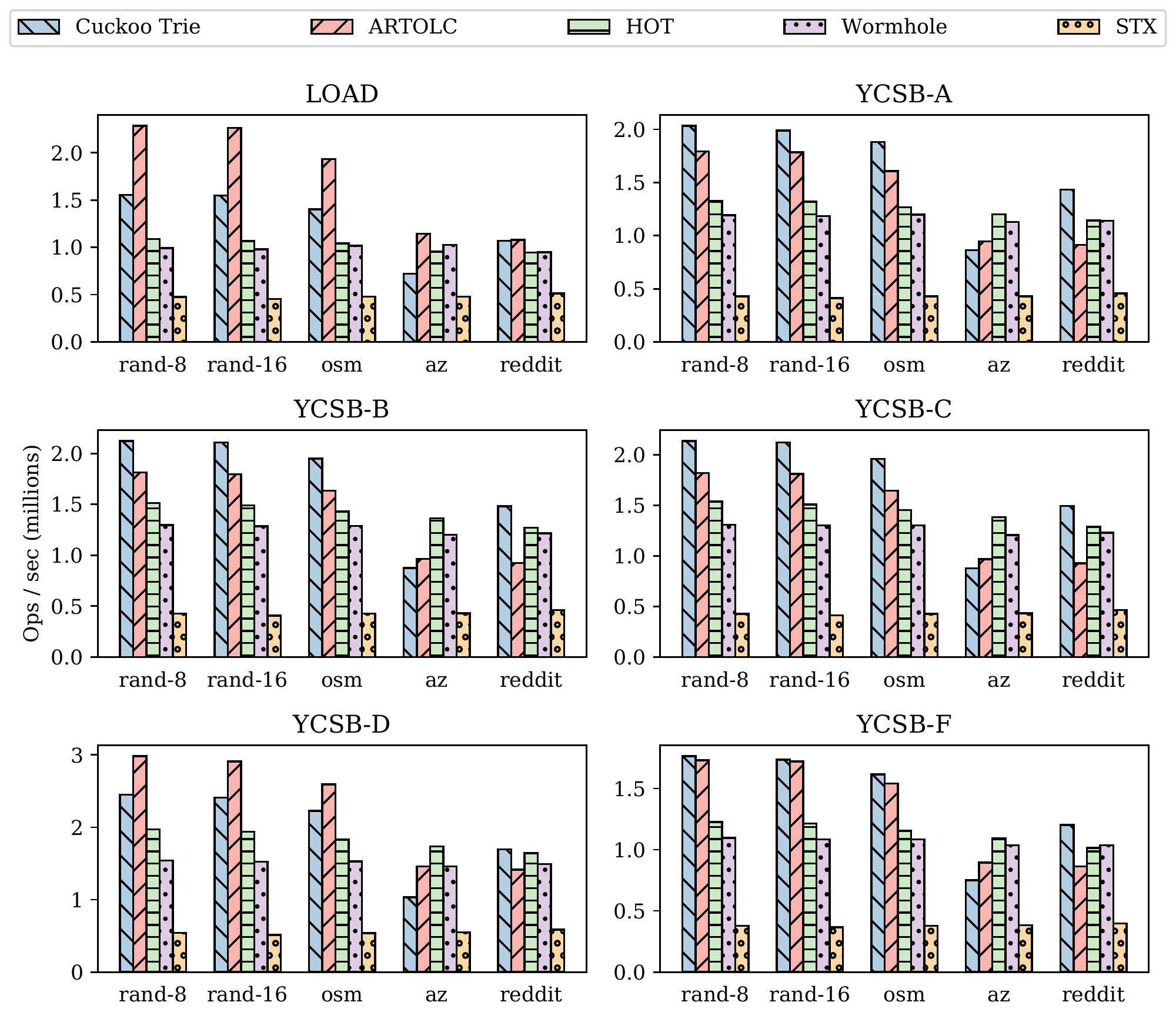}
\vspace{-15pt}
\caption{Single-threaded throughput of YCSB point-operation workloads.}
\label{fig:ycsb_point_1t}
\end{figure*}

First, the Cuckoo Trie is slower on dataset \emph{az}, due to the long common prefixes of its keys. Although these prefixes
are path compressed, this does not save Cuckoo Trie DRAM accesses (\cref{sec:justify-unique}). In contrast, other tries leverage
path compression to save serial accesses (jump down the path). The Cuckoo Trie search thus performs more non-overlapped DRAM accesses
than them.

Second, ARTOLC has high insertion throughput. This is due to the fact that it maintains only a trie, without additional structure (such as a linked list) to accelerate range queries.%
\footnote{To support this explanation, we verified that a Cuckoo Trie that does not maintain the linked list achieves within 5\% of ARTOLC's insert throughput on the \emph{rand-8}, \emph{rand-16} and \emph{osm} datasets.}
The trade-off, however, in ARTOLC's simple structure is that it is not memory efficient (\cref{sec:memory}).

Third, the lookup operations in workload YCSB-D focus on recently-inserted keys. This means that many memory accesses are served from the cache, and therefore the throughput is mostly dependent on the number of instructions executed, not the number of memory accesses. Being much simpler than the other indexes, ARTOLC executes fewer instructions and achieves high throughput in YCSB-D.

\subsection{Scalability to dataset size}

Tree structures usually become slower to search as the dataset size increases, since the tree grows deeper and searches have to traverse longer paths to reach leaves. Moreover, larger trees cannot fully reside in the CPU caches, so searches incur more DRAM accesses and slow down even more.

\begin{figure}
\vspace{-10pt}
\includegraphics[width=\linewidth]{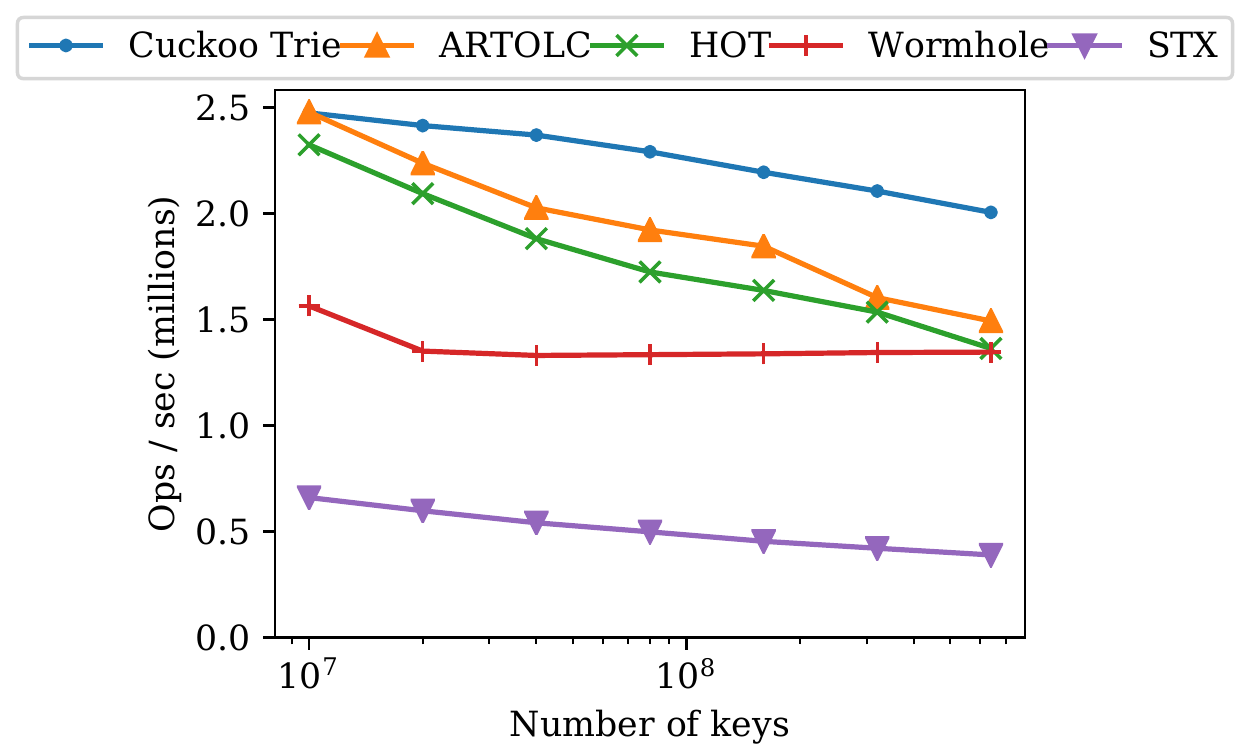}
\vspace{-20pt}
\caption{Single-threaded lookup throughput as a function of dataset size (random 8 byte keys).}
\label{fig:lookup_by_size}
\vspace{-20pt}
\end{figure}

\Cref{fig:lookup_by_size} quantifies this slowdown. It shows the single-threaded lookup throughput (measured over 10\,M random lookups)
obtained after loading the index with datasets of different sizes, from 10\,M to 640\,M random 8-byte keys.

As the dataset grows, the advantage of the Cuckoo Trie over the other indexes becomes more significant. When the dataset size increases from 10\,M to 640\,M keys, the lookup throughput of HOT, ARTOLC, and STX decreases $1.7\times$, $1.66\times$ and $1.69\times$, respectively, while the Cuckoo Trie becomes only $1.23\times$ slower.  While these indexes are all tree-shaped, the Cuckoo Trie's MLP enables reading multiple levels in parallel; for the others, each level implies an additional serial access.

Wormhole's algorithm allows it to achieve nearly constant throughput over the whole range of dataset sizes. However, it is much slower than the Cuckoo Trie for all these sizes.

\subsection{Range iteration performance}

To benchmark range queries, we use the scan-heavy YCSB-E workload (\cref{table:ycsb}). We measure 50\,M YCSB-E operations after loading the dataset, because we observed Wormhole's scan throughput to be low initially (after loading) and taking a few million operations to stabilize.

\begin{figure*}
\begin{subfigure}{\columnwidth}
\includegraphics[width=\columnwidth]{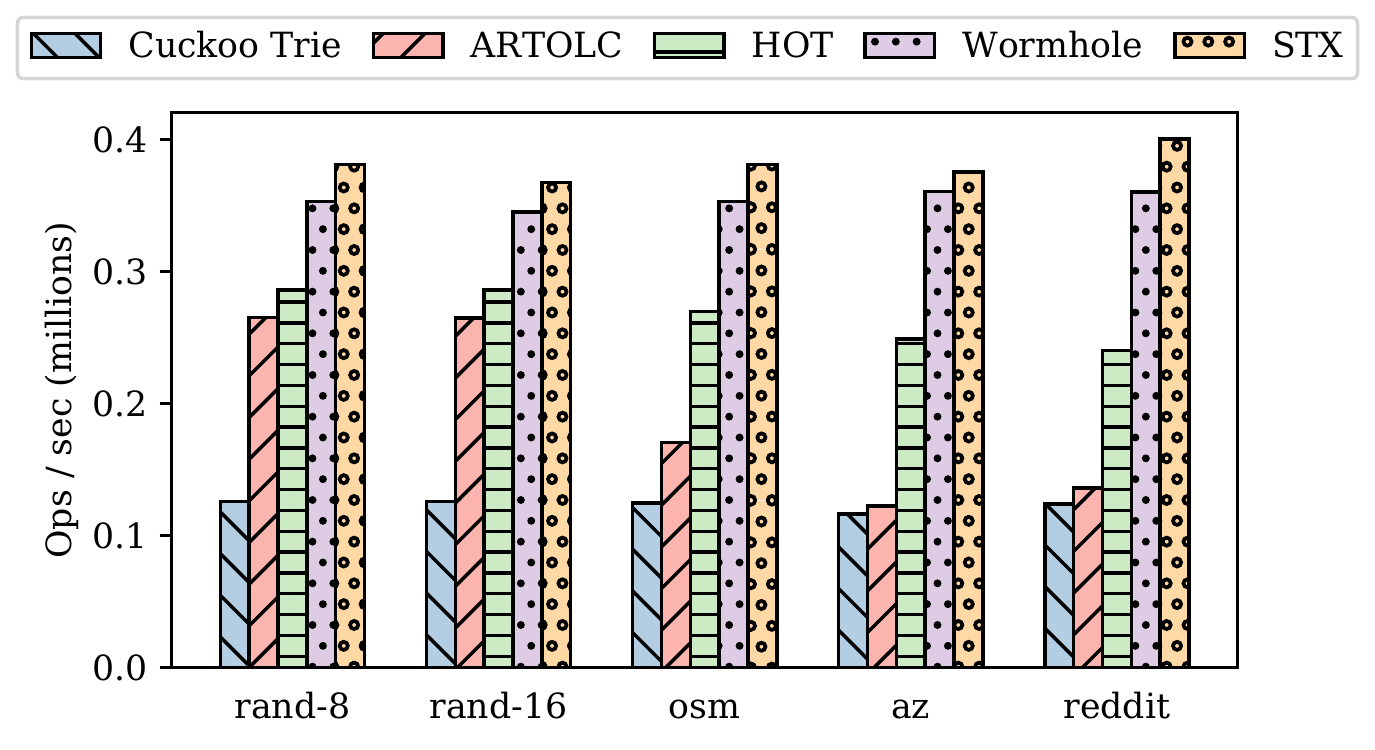}
\vspace{-10pt}
\caption{Single-threaded.}
\end{subfigure}
\begin{subfigure}{\columnwidth}
\includegraphics[width=\columnwidth]{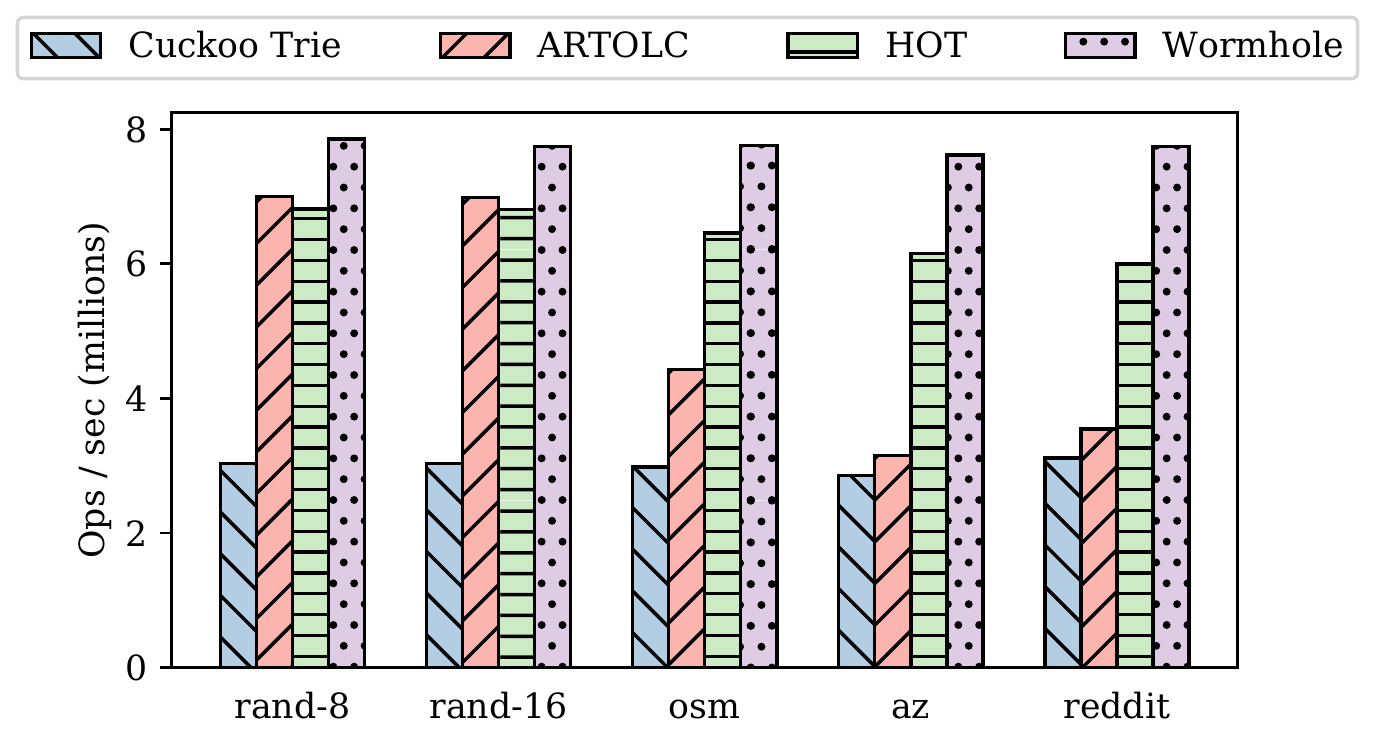}
\vspace{-10pt}
\caption{Multithreaded (28 cores).}
\end{subfigure}
\vspace{-5pt}
\caption{Throughput of scan-heavy YCSB-E workload.}
\vspace{-10pt}
\label{fig:ycsb_e}
\end{figure*}

\Cref{fig:ycsb_e} shows the results. The Cuckoo Trie's scan throughput is lower than that of the other indexes. Here, the Cuckoo Trie's approach of overlapping the serial DRAM accesses of leaf linked list traversal with the client system's work (\cref{sec:range-iters}) is not effective, since the workload performs almost only scans---but without using their results. As we shall see in~\cref{sec:system-benchamrk}, however, when the Cuckoo Trie is part of a real system, these accesses are overlapped with the work done by the system for each element, and so do not affect performance.

\subsection{Memory usage}
\label{sec:memory}

When comparing index memory usage, we have to account for the fact that they expect key-value pairs in different formats that require different amounts of memory. For example, HOT expects null-terminated keys, while Wormhole requires the key length to be stored along with the key as a 32-bit number. To control for these differences, we report index memory \emph{overhead}. That is, the memory required by the index itself, including the pointers to the key-value pairs but excluding the key-value pairs themselves.

As our current Cuckoo Trie implementation does not support automatic resizing,
we measure the memory usage for the hash table with a load factor of 85\%---the load factor used in all our experiments---which we denote $M$. We also estimate the effect that automatic resizing would have, which is relevant for cases where the dataset size is not known
in advance. Assuming that whenever the hash table becomes full its size is increased by a factor of $K$, the expected table size reached for a dataset would be $\frac{1+K}{2} M$. We report this estimate, for $K=2$, as the ``resize'' bar in the figure.

\begin{figure}
\includegraphics[width=\linewidth]{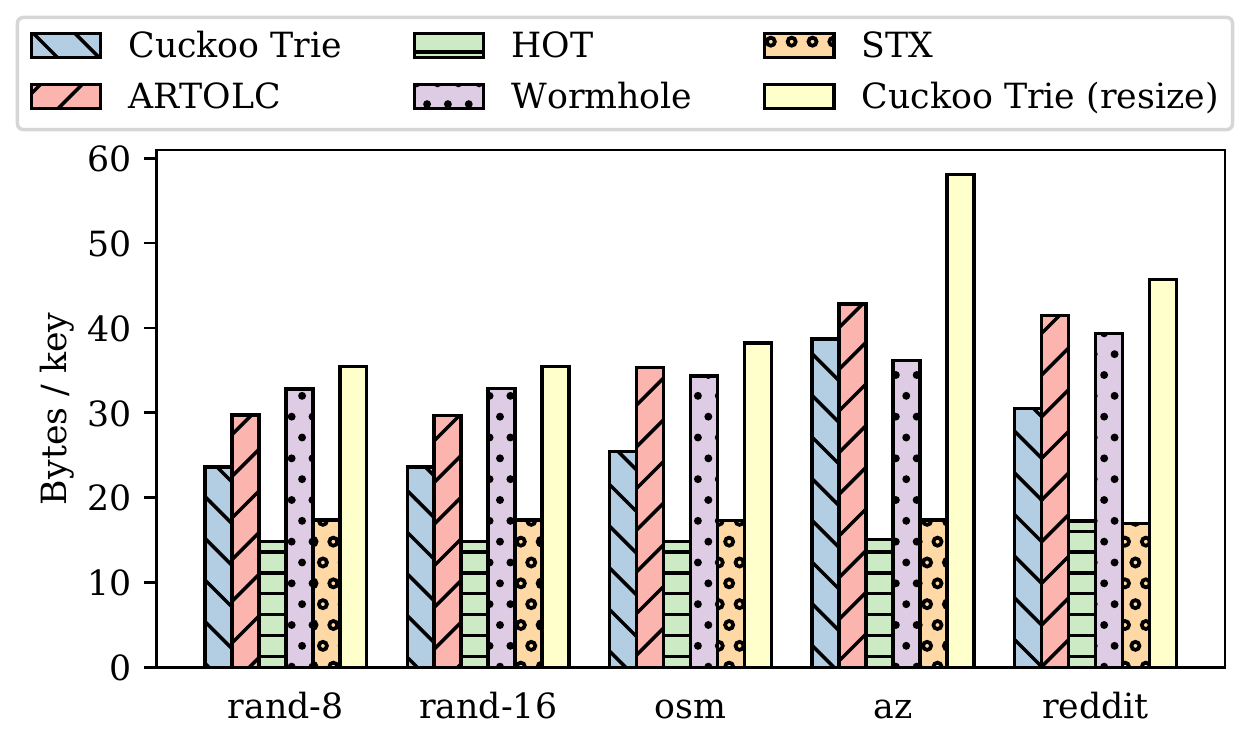}
\vspace{-25pt}
\caption{Memory consumption of the indexes.}
\label{fig:memory}
\vspace{-10pt}
\end{figure}

\Cref{fig:memory} shows the results. For all datasets, the Cuckoo Trie's memory usage is lower than that of ARTOLC and Wormhole by as much as 28\% and higher than that of HOT and STX by as much as 156\%. Even with the effect of automatic resizing, the memory usage is comparable to that of ARTOLC and Wormhole on most datasets. The takeaway is that \emph{the Cuckoo Trie balances memory efficiency with its MLP-related performance.}

\subsection{Memory bandwidth usage} \label{sec:eval:membw}

We evaluate the extent to which the Cuckoo Trie may be bottlenecked by hardware limits such as available memory bandwidth.
We use a workload that maximizes the Cuckoo Trie's memory bandwidth needs: using all 28 cores of the system for a multithreaded YCSB-C (lookup only) workload with the \emph{rand-8} dataset. Because the index size alone exceeds aggregated LLC capacity by more than $100\times$, most Cuckoo Trie memory accesses are satisfied from DRAM.

We compare the Cuckoo Trie's DRAM and UPI bandwidth usage to the hardware's specified limits: $\approx 256$\,GB/sec for DRAM bandwidth (2 $\times$ six DDR4-2666 DRAM channels) and $\approx 93$\,GB/sec for UPI bandwidth (two full-duplex 23.3\,GB/sec links). The specified DRAM bandwidth, however, is for streaming sequential DRAM accesses, whereas the Cuckoo Trie performs random accesses. We therefore also compare the Cuckoo Trie's bandwidth usage to the system's achievable bandwidth for random DRAM reads as measured by Intel's \texttt{mlc} tool~\cite{mlc}.

\Cref{table:membw} shows the results. Even in this worst-case (with respect to bandwidth demands) scenario, the Cuckoo Trie's DRAM bandwidth needs are $3.6\times$ and $2.15\times$ lower than the specified DRAM bandwidth limit and the achievable random access DRAM bandwidth, respectively.
Similarly, the Cuckoo Trie does not fully utilize the NUMA interconnect, though its usage is closer to the hardware limit. The reason is that UPI traffic includes cache coherence traffic in addition to data transfers. Using the \texttt{pcm} tool~\cite{pcm}, we find that UPI data transfers account for 35.6\,GB/sec of traffic---consistent with each node fetching half of its data from remote DRAM, as expected with NUMA-interleaved memory allocations.

\begin{table}
\centering
\small
\begin{tabular}{>{\arraybackslash}m{1.5cm}|
				>{\centering\arraybackslash}m{2.5cm}
				>{\centering\arraybackslash}m{1.5cm}
				>{\centering\arraybackslash}m{1.5cm}}
\toprule
\textbf{Resource} & \textbf{Bandwidth usage}                        & \textbf{\% of specified maximum} & \textbf{\% of rand-read maximum} \\
\midrule
DRAM              & 71.24\,GB/sec (= 1113\,M accesss/sec) & 27.8\%                           & 46.3\% \\
UPI               & 61\,GB/sec                            & 65.5\%                           & 68\%   \\
\bottomrule
\end{tabular}
\caption{DRAM and UPI bandwidth usage of multithreaded (28 cores) Cuckoo Trie on the YCSB-C workload with the \emph{rand-8} dataset.}
\label{table:membw}
\vspace{-25pt}
\end{table}

\subsection{Comparison with MlpIndex}

\begin{figure}
	\includegraphics[width=0.9\linewidth]{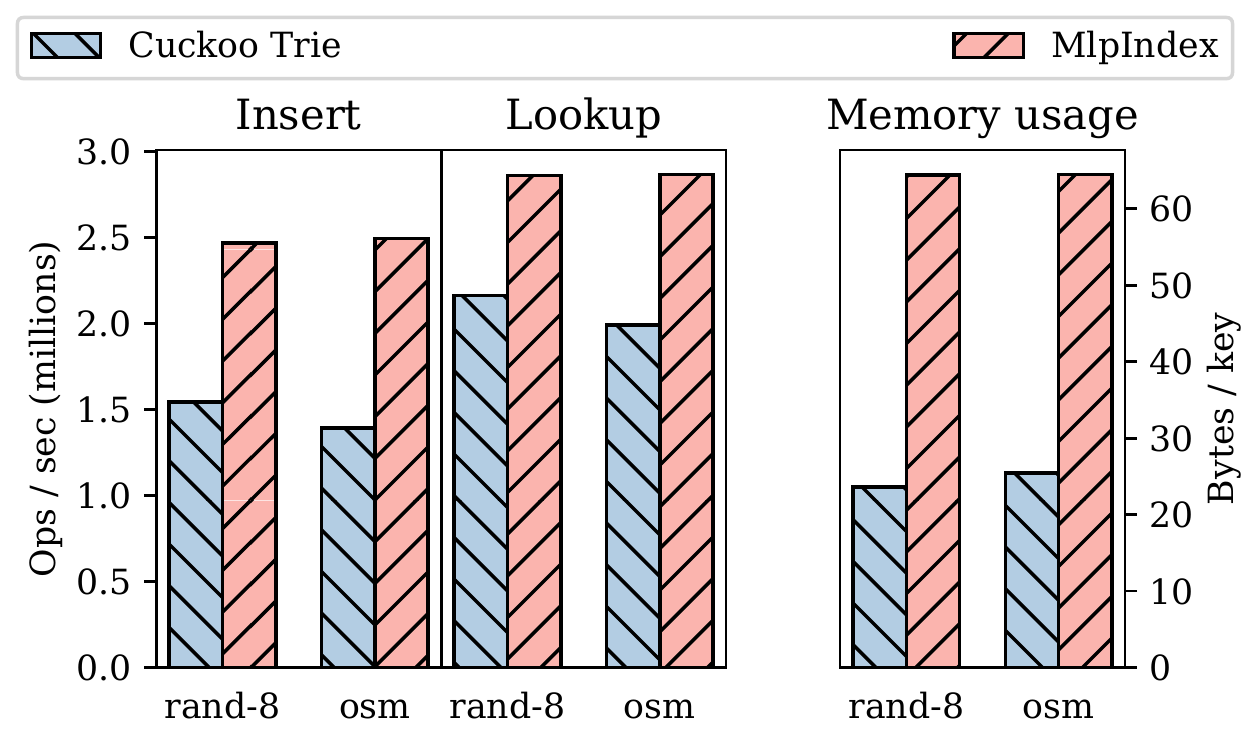}
    \vspace{-15pt}
	\caption{Comparison of the Cuckoo Trie and MlpIndex.}
	\label{fig:mlpindex}
    \vspace{-10pt}
\end{figure}

MlpIndex~\cite{mlp-index} is another MLP-aware index. It is also based on a hashed trie representation, but its simple representation only supports 8-byte keys, and it does not support range scans and multithreading.
\Cref{fig:mlpindex} compares the insert throughput (LOAD workload), lookup throughput (10\,M random lookups), and memory overhead of the CuckooTrie and MlpIndex on the two 8-byte-key datasets. Each index is initialized to the minimal size that allows loading the dataset.

MlpIndex outperforms the Cuckoo Trie by 30\%-80\% in the insertion and lookup benchmarks, because it exploits the fact that keys
are 8-bytes to embed them directly in the leaves, thus saving one pointer dereference. However, because MlpIndex has large nodes and no key elimination, its memory overhead is $\approx 3\times$ higher than the Cuckoo Trie's.

\subsection{Full-system benchmark}
\label{sec:system-benchamrk}

We evaluate the Cuckoo Trie as part of a real system. We use Redis~\cite{redis}, a popular in-memory data store. Redis offers a \emph{sorted set} that provides the ordered index API. Redis' implementation uses a hash table for point lookups and a skip list for range scans, with keys inserted into both structures.

We extend Redis to support using the benchmarked indexes as its sorted set data type implementation.
We evaluate the YCSB workloads for each implementation. To eliminate network and storage delays, the client and server are run on different cores of the same processor, and saving the database to disk was disabled in the Redis configuration. The server is restarted after each workload. We use 4 client threads, which was experimentally determined to give the best performance.

\begin{figure*}
\includegraphics[width=0.8\linewidth]{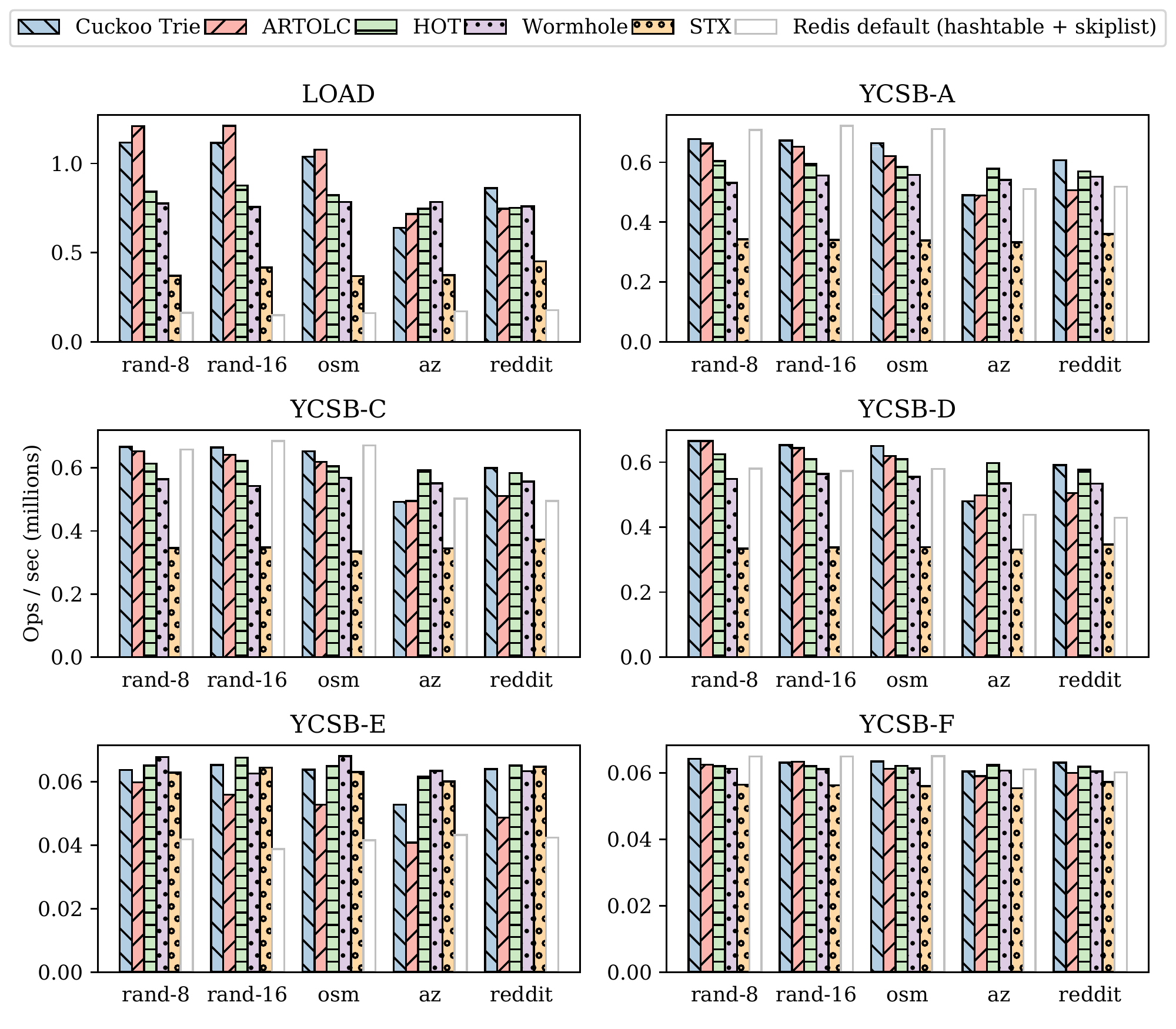}
\vspace{-10pt}
\caption{Redis throughput with different sorted-set implementations. (YCSB-B omitted as it is nearly identical to YCSB-C.)}
\label{fig:ycsb_redis}
\end{figure*}

\Cref{fig:ycsb_redis} shows the results. On YCSB workloads A--D, which contain both inserts and lookups, the Cuckoo Trie gives the best Redis performance on all datasets except \emph{az}. On YCSB-C, the lookup-only workload, the Cuckoo Trie performs as fast as the default Redis sorted set. This suggests that replacing the skip list in the default Redis implementation with the Cuckoo Trie would make the hash table unnecessary, thus saving memory without affecting performance.

On the scan-heavy YCSB-E, the throughput of the Cuckoo Trie is comparable to that of the best performing indexes (again, except for \emph{az}). This shows that when the Cuckoo Trie is part of a system, the latency associated with linked list traversal gets overlapped with other work done by the system, and does not have a significant impact.

\section{Related work} \label{sec:related}

There is much work on in-memory indexes. We focus on the aspects most relevant to our work:

\noindentparagraph{Optimizing index memory access.}
Prior optimizations include laying index data sequentially in memory~\cite{csb,css-tree,hydra-list}; prefetching wide B-Tree nodes~\cite{PrefetchingBtree,masstree}; increasing trie symbol size~\cite{art,masstree}; and dynamically varying the symbol size~\cite{hot}.
While these optimizations lower the number or latency of memory accesses, the accesses are still \emph{inherently sequential}.

\noindentparagraph{MLP.}
To our knowledge, the only prior index that uses MLP is MlpIndex~\cite{mlp-index}. It also uses a hashed trie representation, but stores full keys in hash table entries, whereas we use key elimination. MlpIndex thus only supports fixed-length keys and has a large memory footprint.
It also does not support range scans and multithreading.
Cimple~\cite{cimple} introduces MLP by interleaving several queries to a standard data structure, but this can only accelerate \emph{batches} of \emph{independent} queries.

\noindentparagraph{Hashed representation.}
Hashed representations are used by X-Fast tries~\cite{xfast} and Wormhole~\cite{wormhole}. Both, however, use serial binary search to determine the depth of a key in the trie, whereas the Cuckoo Trie tests multiple depths in parallel.

\noindentparagraph{Key elimination.}
Key elimination is inspired by the Bonsai dictionary~\cite{bonsai,m-bonsai}, which also stores key prefixes in a hash table using a constant amount of information per node. In contrast to our design, Bonsai is inherently sequential: locating a node requires reading its parent first.

\section{Conclusion}

We presented the Cuckoo Trie, a fast, memory-efficient ordered index designed with MLP as a primary consideration.
MLP makes the Cuckoo Trie faster than current indexes, while using a novel key-eliminating hash table design makes it memory efficient.
Our evaluation demonstrated that the Cuckoo Trie outperforms state-of-the-art indexes, typically with a smaller or comparable
memory footprint.

This work highlights the potential benefits of making MLP a first-order design consideration.
For the last two decades, DRAM capacity has been doubling but its latency has stagnanted~\cite{dram-trends}.
Assuming these trends continue, MLP-awareness will become even more important in the future.

The Cuckoo Trie code is available at \url{https://github.com/cuckoo-trie/cuckoo-trie-code}.

\vspace{-5pt}
\begin{acks}
This work was funded in part by ISF under grant 2005/17, Blavatnik ICRC at TAU, and the Blavatnik Family Foundation.
We thank the reviewers and the paper's shepherd, Dushyanth Narayanan, for their feedback.
\end{acks}

\bibliographystyle{ACM-Reference-Format}
\bibliography{paper}

\end{document}